\documentclass[12pt]{article}

\usepackage{latexsym,amsmath,amscd,amssymb,graphics}
\usepackage{enumerate}

\usepackage{graphicx}

\usepackage[square,authoryear]{natbib}
\usepackage{url}

\usepackage[all]{xy}

\makeatletter

\@addtoreset{figure}{section}
\def\thefigure{\thesection.\@arabic\c@figure}
\def\fps@figure{h, t}
\@addtoreset{table}{bsection}
\def\thetable{\thesection.\@arabic\c@table}
\def\fps@table{h, t}
\@addtoreset{equation}{section}

\makeatother

\begin{document}

\newtheorem{theorem}{Theorem}[section]
\newtheorem{definition}[theorem]{Definition}
\newtheorem{lemma}[theorem]{Lemma}
\newtheorem{remark}[theorem]{Remark}
\newtheorem{proposition}[theorem]{Proposition}
\newtheorem{corollary}[theorem]{Corollary}
\newtheorem{example}[theorem]{Example}

\def\below#1#2{\mathrel{\mathop{#1}\limits_{#2}}}



\title{Reduced Lagrangian and Hamiltonian formulations of Euler-Yang-Mills fluids}
\author{Fran\c{c}ois Gay-Balmaz$^{1}$ and Tudor S. Ratiu$^{1}$}
\addtocounter{footnote}{1} \footnotetext{Section de
Math\'ematiques and Bernoulli Center, \'Ecole Polytechnique F\'ed\'erale de
Lausanne. CH--1015 Lausanne. Switzerland. Partially supported by Swiss NSF
grant
200021-109111.
\texttt{Francois.Gay-Balmaz@epfl.ch, Tudor.Ratiu@epfl.ch}
\addtocounter{footnote}{1} }


\date{ }
\maketitle

\makeatother
\maketitle


\noindent \textbf{AMS Classification:} 37K65, 53C80, 70S15, 53D17, 76W05

\noindent \textbf{Keywords:} Euler-Poincar\'e equations, Lie-Poisson equations,
Euler-Yang-Mills equations, automorphism group, gauge group, reduction,
Kaluza-Klein metric, Poisson bracket

\begin{abstract} The Lagrangian and Hamiltonian structures for an ideal
gauge-charged fluid are determined. Using a Kaluza-Klein point of view, the
equations of motion are obtained by Lagrangian and Poisson reductions
associated
to the automorphism group of a principal bundle. As a consequence of the
Lagrangian approach, a Kelvin-Noether theorem is obtained. The Hamiltonian
formulation determines a non-canonical Poisson bracket associated to these
equations.
\end{abstract}

\section{Introduction}\label{introduction}

The equations of motion of an \textbf{ideal incompressible fluid} on an
oriented
Riemannian manifold $(M,g)$ are given by the Euler equations
\begin{equation}\label{Euler}
\frac{\partial v}{\partial t}+\nabla_vv=-\operatorname{grad}p,
\end{equation}
where the Eulerian velocity $v$ is a divergence free vector field, $p$ is the
pressure, and $\nabla$ is the Levi-Civita covariant derivative associated to
$g$. \cite{Arnold1966} has shown
that equations \eqref{Euler} are formally the spatial
representation of the geodesic spray on the volume-preserving
diffeomorphism group $\mathcal{D}_\mu(M)$ of $M$ with respect to
the $L^2$ Riemannian metric, where $\mu$ is the Riemannian volume form on $M $.
See also \cite{AbMa1978}, \S5.5.8,
for a quick exposition of this fact. \cite{EbMa1970} give the
analytic formulation and many rigorous results concerning the
Euler and Navier-Stokes equations derived from this geometric
point of view. From the Hamiltonian perspective, equations
\eqref{Euler} are the  Lie-Poisson equations on the Lie algebra
$\mathfrak{X}_{\rm div}(M)$  of $\mathcal{D}_\mu(M)$, consisting of
divergence free vector fields (\cite{MaWe1983}). Here, 
$\mathfrak{X}_{\rm div}(M)$ is identified with its dual by the weak
$L^2$-pairing.

In \cite{MaWeRaScSp1983} and \cite{MaRaWe1984}, this approach is generalized to
the case of the motion of an \textbf{ideal compressible adiabatic fluid}
\begin{equation}\label{ICAF}
\left\lbrace
\begin{array}{ll}
\vspace{0.2cm}\displaystyle\frac{\partial v}{\partial
t}+\nabla_vv=\frac{1}{\rho}\operatorname{grad}p,\\
\vspace{0.2cm}\displaystyle\frac{\partial \rho}{\partial
t}+\operatorname{div}(\rho v)=0,\\
\displaystyle\frac{\partial s}{\partial t}+\textbf{d}s(v)=0,
\end{array} \right.
\end{equation}
where $\rho$ is the mass density, $s$ is the specific entropy, and $p$ is
the pressure. In this case, the configuration space is the full diffeomorphism
group $\mathcal{D}(M)$ and equations \eqref{ICAF} are obtained via Lie-Poisson
reduction for semidirect products. The Euler-Poincar\'e approach is given
in \cite{HoMaRa1998}.

In this work we generalize the two previous procedures to the case of a
\textbf{classical charged ideal fluid}. More precisely, using a Kaluza-Klein
point of view, we obtain the equations of motion by Lagrange-Poincar\'e (see
\cite{CeMaRa2001}) and Poisson reduction by a symmetry group (see, e.g.
\cite{MaRa1994}, \S10.5). We consider on $M$ a $G$-principal
bundle $P\rightarrow M$ and enlarge the configuration space from the group
of
diffeomorphisms of $ M $ to the product of the group of automorphisms of $P$ with the
field variables.

If $G=S^1$ we recover the \textbf{Euler-Maxwell equations} describing the motion
of
an electrically charged fluid
\begin{equation}\label{EulerMaxwell}
\left\lbrace
\begin{array}{ll}
\vspace{0.2cm}\displaystyle\frac{\partial v}{\partial
t}+\nabla_vv=\frac{q}{m}(\textbf{E}+v\times
\textbf{B})-\frac{1}{\rho}\operatorname{grad}p,\\
\vspace{0.2cm}\displaystyle\frac{\partial \rho}{\partial
t}+\operatorname{div}(\rho v)=0,\\
\vspace{0.2cm}\displaystyle\frac{\partial s}{\partial
t}+\textbf{d}s(v)=0,\\
\vspace{0.2cm}\displaystyle\frac{\partial \textbf{E}}{\partial
t}=\operatorname{curl}\textbf{B}-\frac{q}{m}\rho v,\quad\frac{\partial
\textbf{B}}{\partial t}=-\operatorname{curl}\textbf{E},\\
\displaystyle\operatorname{div}\textbf{E}=\frac{q}{m}\rho,\quad\operatorname{div}\textbf{B}=0,
\end{array} \right.
\end{equation}
where $v$ is the Eulerian velocity, $\textbf{E}$ is the electric field,
$\textbf{B}$ is the magnetic field, $m$ is the mass of the charged fluid
particles,  and the constant $q$ is the electric charge of the particles. The
Hamiltonian structure of the incompressible Euler-Maxwell equations is 
already presented in \cite{MaWeRaScSp1983}.
  
Returning to the general case of a $G$-principal bundle, we will show that the
Lagrange-Poincar\'e and Poisson reduction methods lead to the equations for an
ideal compressible adiabatic fluid carrying a gauge-charge, as given in
\cite{GiHoKu1983}. We call these equations the 
$\textbf{Euler-Yang-Mills}$ equations. The physically relevant examples are
obtained for $G=\operatorname{SU}(2)$ or $G=\operatorname{SU}(3)$ in which case
the associated fluid motion goes also under the name of chromohydrodynamics. For
a Lagrangian description of the Euler-Yang-Mills equations and the associated
variational principle formulated in local coordinates both in the
non-relativistic and relativistic versions, see \cite{BiJaLiNaPi2003} and
\cite{JaNaPiPo2004}. In these papers the variations are constrained according to
the general Euler-Poincar\'e variational principle for field theories presented
in \cite{CaRaSh2000}.

The physical interpretation of the equations obtained by the methods given in
this paper is the
following. The evolution of the fluid particles as well as of the gauge-charge
density of the fluid is given by a curve $\psi_t$ in the automorphism group of a principal bundle $P \rightarrow  M$. In fact, $\psi_t$ is the flow of a time-dependent vector field $U_t$ on $P$. This vector field induces a time-dependent vector
field $v_t$ on $M$, which represents the \textbf{Eulerian velocity of the
fluid}. Given the evolution of the Yang-Mills fields potential and of the mass
density, the vector field $U_t$ induces also a Lie algebra valued and
time-dependent function which represents the \textbf{gauge-charge density of
the fluid}. Note the analogy with the classical Kaluza-Klein construction
appearing in the formulation of the equations of motion for a charged
particle in a Yang-Mills field. See also \cite{BaMaMu2006} who generalize the
Eulerian fluid velocity vector to include a non-Abelian, or gauge, index.

The paper is organized as follows. To fix notations and conventions, we
summarize in \S\ref{Connections} some basic facts about principal bundles,
connections, automorphisms, and gauge groups. The Hamiltonian and Lagrangian
formulations of the Maxwell equations are recalled in \S\ref{Fields} and
generalized to the case of the Yang-Mills
fields equations. The Lagrangian formulation of the motion of a charged
classical particle in a Yang-Mills field, that is, the Wong equations, are
presented in \S\ref{Particles}. In \S\ref{Lagrangian} it is shown that the
compressible and incompressible Euler-Yang-Mills equations consist of coupled
equations. These are the Euler-Poincar\'e equations of a semidirect product
(associated to the automorphism group of a principal bundle) for the fluid and
charge variables and the Yang-Mills equations for the vector potential (that is,
the connection) and the ``electric part" of the Yang-Mills field. The
Hamiltonian counterpart of this result is presented in \S\ref{Hamiltonian}: one
obtains coupled equations consisting  of Lie-Poisson equations on the same dual
for the fluid and charge variables together with the Yang-Mills equations.
Formally the Gauss equation relating the gauge-charge and the ``electric
part" of the Yang-Mills field is missing from this system. However, it is
obtained by conservation of the momentum map associated to the invariance under gauge transformations. We naturally obtain the non-canonical Poisson bracket associated to the Hamiltonian formulation of the Euler-Yang-Mills equations. By applying the general process of reduction
by stages, we recover some already known results about the Euler-Maxwell
equations. We also show that the two different Poisson brackets derived in
\cite{GiHoKu1983} and in \cite{MaWeRaScSp1983} are in fact obtained by Poisson
reduction, at different stages, of the same canonical Poisson structure.
Finally, in \S\ref{section:KN} we present a Kelvin-Noether Theorem for the
Euler-Yang-Mills equations.

\section{Connections, automorphisms, and gauge
transformations}\label{Connections}

In this section we recall basic notions related to principal bundles. We also
introduce notations and conventions that will be used throughout the paper. 

\medskip

\subsection{Principal and adjoint bundles}  Consider a smooth free and proper
right action
\[
\Phi : G\times P\rightarrow P,\;\;(g,p)\mapsto\Phi_g(p)
\]
of a Lie group $G$ on a manifold $P$. Thus we get   the principal bundle
\[
\pi :P\rightarrow M:=P/G,
\]
where $M$ is endowed with the unique manifold structure for which $\pi$ is a
submersion.

To any element $\xi$ in the Lie algebra $\mathfrak{g}$ of $G$ there corresponds
a
vector field $\xi_P$ on $P$, called the \textbf{infinitesimal generator},
defined by
\[
\xi_P(p):=\frac{d}{dt}\Big{|}_{t=0}\Phi_{\operatorname{exp}(t\xi)}(p).
\]
At any $p\in P$, these vector fields generate the \textbf{vertical subspace}
\[
V_pP:=\{\xi_P(p)\mid \xi\in\mathfrak{g}\}=\operatorname{ker}(T_p\pi).
\]

Recall that the \textbf{adjoint vector bundle} is
\[
\operatorname{Ad}P:=P\times_G\mathfrak{g}\rightarrow M,
\]
where the quotient is taken relative to the right action $(g,(p,\xi))\mapsto
(\Phi_g(p), \operatorname{Ad}_{g^{-1}}(\xi))$. The elements of
$\operatorname{Ad}P$ are denoted by $[p,\xi]_G$, for $(p,\xi)\in
P\times\mathfrak{g}$. There is a Lie bracket operation $[ \cdot , \cdot]_x$ on
each fiber  $\left(\operatorname{Ad}P\right)_x$ depending smoothly on $ x \in  M
$; it is defined by
\[
\left[ [p, \xi]_G, [p, \eta]_G \right]_x : = [p, [ \xi, \eta]]_G
\]
for $[p, \xi]_G, [p, \eta]_G \in \operatorname{Ad}P $, $\pi(p) = x$.

\subsection{Exterior forms on adjoint bundles} Consider the space
$\Omega^k(P,\mathfrak{g})$ of $\mathfrak{g}$-valued
$k$-forms
on $P$ and let $\overline{\Omega^k}(P,\mathfrak{g})$ be the subspace of
$\Omega^k(P,\mathfrak{g})$ consisting of $\mathfrak{g}$-valued $k$-forms
$\omega$ such that:
\begin{enumerate}
\item[(1)] $\Phi_g^*\omega=\operatorname{Ad}_{g^{-1}}\circ\,\omega$,
\item[(2)] if one of $u_1,...,u_k\in T_pP$ is vertical then
$\omega(u_1,...,u_k)=0$.
\end{enumerate}

The real vector space $\overline{\Omega^k}(P,\mathfrak{g})$ is naturally
isomorphic to $\Omega^k(M,\operatorname{Ad}P)$, the space of
$\operatorname{Ad}P$-valued $k$-forms on $M$. Indeed, to each
$\omega\in\overline{\Omega^k}(P,\mathfrak{g})$
corresponds a $k$-form $\widetilde{\omega}\in\Omega^k(M,\operatorname{Ad}P)$
whose value on $v_1,..,v_k\in T_xM$ is given by
\begin{equation}\label{tilde}
\widetilde{\omega}(x)(v_1,...,v_k):=[p,\omega(p)(u_1,...,u_k)]_G,
\end{equation}
where $p\in P$ is such that $\pi(p)=x$ and $u_i\in T_pP$ are such that
$T_p\pi(u_i)=v_i$. 

To define the inverse of the map $\;\widetilde\; :
\overline{\Omega^k}(P,\mathfrak{g}) \rightarrow \Omega^k(M,\operatorname{Ad}P)$
we introduce first for every $p \in P $ the $\mathbb{R}$-linear map $i_p:
(\operatorname{Ad} P)_x \rightarrow \mathfrak{g}$, $x: = \pi(p) \in M $, by 
\[
i_p\left( [q,\eta]_G\right) :=\xi,\quad\text{where $\xi$ is such that}\quad
[p,\xi]_G=[q,\eta]_G;
\]
in this formula $p, q \in P $ and $ \xi, \eta\in \mathfrak{g}$. Equivalently,
this definition can be restated as
\[
i_p\left( [q,\eta]_G\right) :=\operatorname{Ad}_g \eta,\;\text{where $g \in
G$ is uniquely determined by }\;
q = \Phi_g(p).
\]
Then the definition of the Lie bracket on each fiber $( \operatorname{Ad}P)_x$
of the adjoint bundle $\operatorname{Ad}P$ immediately implies that $i_p:
(\operatorname{Ad}P)_x \rightarrow \mathfrak{g}$ is a Lie algebra isomorphism.
In addition, $i_{\Phi_g(p)}=\operatorname{Ad}_{g^{-1}}\circ i_p$ for every $p
\in P $ and $g \in G $. Using  the maps $i_p$ for every $p \in P $, define the
inverse of  $\;\widetilde\;: \overline{\Omega^k}(P,\mathfrak{g}) \rightarrow
\Omega^k(M,\operatorname{Ad}P)$  by
\begin{equation}\label{tilde_inverse}
\omega(p)(u_1,...,u_k):=i_p\Bigl(\widetilde{\omega}(\pi(p))\bigl(T_p\pi(u_1),...,T_p\pi(u_1)\bigr)\Bigr),
\end{equation}
for any $p \in P $ and $u_1, \ldots, u_k \in T_pP$. The identity
$i_{\Phi_g(p)}=\operatorname{Ad}_{g^{-1}}\circ i_p$
ensures that $\omega\in\overline{\Omega^k}(P,\mathfrak{g})$.

Since $\overline{\Omega^0}(P,\mathfrak{g}) = \mathcal{F}_G(P, \mathfrak{g}) :=
\{f:P\rightarrow\mathfrak{g} \mid
f\circ\Phi_g=\operatorname{Ad}_{g^{-1}}\circ f\}$  and
$\Omega^0(M,\operatorname{Ad}P) = \Gamma(\operatorname{Ad}P)$,  the space of
sections of $\operatorname{Ad}P$, we shall use the notations
$\overline{\Omega^0}(P,\mathfrak{g})=\mathcal{F}_G(P,\mathfrak{g})$ and
$\Omega^0(M,\operatorname{Ad}P)=\Gamma(\operatorname{Ad}P)$ interchangeably. We
have hence $\mathcal{F}_G(P, \mathfrak{g}) \cong \Gamma( \operatorname{Ad} P)$
as Lie algebras, the isomorphism $f \in \mathcal{F}_G(P, \mathfrak{g}) \mapsto
\widetilde{f} \in \Gamma( \operatorname{Ad}P) $ being given by  \eqref{tilde},
that is, $\widetilde{f}(\pi(p)) = [p, f(p)]_G$.

\subsection{Connections and covariant differentials} A \textbf{principal
connection} on $P$ is a $\mathfrak{g}$-valued $1$-form
$\mathcal{A}\in\Omega^1(P,\mathfrak{g})$ such that
\[
\Phi_g^*\mathcal{A}=\operatorname{Ad}_{g^{-1}}\circ\,\mathcal{A}\;\;\text{and}\;\;\mathcal{A}(\xi_P)=\xi.
\]
The set of all connections will be denoted by $\mathcal{C}onn(P)$. It is an
affine space with underlying vector space
$\overline{\Omega^1}(P,\mathfrak{g})$. Recall that a connection induces a
splitting $T_pP=V_pP\oplus H_pP$
of the tangent space into the vertical and \textbf{horizontal subspace} defined
by
\[
H_pP:=\operatorname{ker}(\mathcal{A}(p)).
\]

The \textbf{covariant exterior differential} associated to $\mathcal{A}$ is the
map
$\textbf{d}^\mathcal{A}:\Omega^k(P,\mathfrak{g})\rightarrow\Omega^{k+1}(P,\mathfrak{g})$
defined by
\[
\textbf{d}^\mathcal{A}\omega(p)(u_1,...,u_k)
:=\textbf{d}\,\omega(p)\left(\operatorname{hor}_p(u_1),...,
\operatorname{hor}_p(u_k)\right),
\]
where $\operatorname{hor}_p(u_i)$ is the horizontal part of $u_i
\in T_pP$, $i = 1, \dots, k$. Note that for
$\omega\in\overline{\Omega^k}(P,\mathfrak{g})$ we have
$\textbf{d}^\mathcal{A}\omega\in\overline{\Omega^{k+1}}(P,\mathfrak{g})$.
For $f\in\mathcal{F}_G(P,\mathfrak{g})$ and $\omega \in
\overline{ \Omega^1}(P, \mathfrak{g}) $ we have the formulas
\begin{align*}
\textbf{d}^\mathcal{A}f(p)(u) &=\textbf{d}f(p)(u)+[\mathcal{A}(p)(u),f(p)]
\quad \text{and} \quad \\
\mathbf{d}^{\mathcal{A}} \omega(p) (u,v) &= \mathbf{d} \omega(p)(u,v) + [
\mathcal{A}(p)(u), \omega(p)(v)] - [ \mathcal{A}(p)(v), \omega(p)(u)]
\end{align*}
for any $u, v \in T_pP$.

The \textbf{curvature of the connection} $\mathcal{A}$ is, by definition, the
$2$-form
\[
\mathcal{B}:=\textbf{d}^\mathcal{A}\mathcal{A}\in\overline{\Omega^2}(P,\mathfrak{g}).
\]
The curvature $\mathcal{B}$ verifies the Cartan Structure Equations
and the Bianchi Identity given respectively by
\[
\mathcal{B}(u,v)=\textbf{d}\,\mathcal{A}(u,v)+[\mathcal{A}(u),\mathcal{A}(v)]\quad\text{and}\quad\textbf{d}^\mathcal{A}\mathcal{B}=0.
\]
The following lemma will be useful for future computations.

\begin{lemma}\label{d^Ad^A} Let $\mathcal{A}\in\mathcal{C}onn(P)$,
$\mathcal{B}$
its curvature, and $f\in\mathcal{F}_G(P,\mathfrak{g})$. Then
\[
{\bf d}^\mathcal{A}{\bf d}^\mathcal{A}f(u,v)=[\mathcal{B}(u,v),f].
\]
\end{lemma}
\textbf{Proof.} For any $U,V\in\mathfrak{X}(P)$ we have
\begin{align*}\textbf{d}^\mathcal{A}(\textbf{d}^\mathcal{A}&f)(U,V)=\textbf{d}(\textbf{d}^\mathcal{A}f)(U,V)+[\mathcal{A}(U),\textbf{d}^\mathcal{A}f(V)]-[\mathcal{A}(V),\textbf{d}^\mathcal{A}f(U)]\\
&=\textbf{d}\left(\textbf{d}f+[\mathcal{A},f]\right)(U,V)+[\mathcal{A}(U),\textbf{d}f(V)+[\mathcal{A}(V),f]]\\
&\qquad-[\mathcal{A}(V),\textbf{d}f(U)+[\mathcal{A}(U),f]]\\
&=0+\textbf{d}([\mathcal{A}(V),f])(U)-\textbf{d}([\mathcal{A}(U),f])(V)-[\mathcal{A}([U,V]),f]\\
&\qquad
+[\mathcal{A}(U),\textbf{d}f(V)]+[\mathcal{A}(U),[\mathcal{A}(V),f]]-[\mathcal{A}(V),\textbf{d}f(U)] \\
& \qquad -[\mathcal{A}(V),[\mathcal{A}(U),f]]\\
&=[\textbf{d}(\mathcal{A}(V))(U),f]-[\textbf{d}(\mathcal{A}(U))(V),f]-[\mathcal{A}([U,V]),f]\\
&\qquad
+[\mathcal{A}(U),[\mathcal{A}(V),f]]-[\mathcal{A}(V),[\mathcal{A}(U),f]]\\
&=[\textbf{d}\mathcal{A}(U,V),f]+[[\mathcal{A}(U),\mathcal{A}(V)],f]\\
&=[\mathcal{B}(U,V),f].\qquad\blacksquare
\end{align*}

Recall also that a principal connection $\mathcal{A}$ on $P$ induces an affine
connection and a covariant derivative, denoted respectively by
$\nabla^\mathcal{A}$ and $\frac{D^\mathcal{A}}{dt}$, on the vector bundles
$\operatorname{Ad}P\rightarrow M$ and $(\operatorname{Ad}P)^*\rightarrow M$
(see
e.g. \cite{KoNo1963} or \cite{CeMaRa2001}).
\medskip

Given a Riemannian metric $g$ on $M$ and a connection $\mathcal{A}$ on $P$, we
can
define the \textbf{covariant codifferential}
\[
\delta^\mathcal{A} :
\overline{\Omega^k}(P,\mathfrak{g})\rightarrow
\overline{\Omega^{k-1}}(P,\mathfrak{g});
\]
see, e.g., Definition 4.2.8 in \cite{Bleecker1981}.

\subsection{Bundle metrics}  We assume throughout this paper that the Lie
algebra $\mathfrak{g}$ has a distinguished inner product $\gamma$ satisfying
\[
\gamma(\operatorname{Ad}_g\xi,\operatorname{Ad}_g\eta)=\gamma(\xi,\eta),\;\;\text{for
all}\;\;g\in G\;\; \text{and all}\;\; \xi, \eta \in \mathfrak{g}.
\]
Such an inner product is said to be $\operatorname{Ad}$-invariant and satisfies
the relation
\begin{equation}\label{Ad-invariance}
\gamma([\zeta,\xi],\eta)+\gamma(\xi,[\zeta,\eta])=0 \;\;\text{for all}\;\; \xi,
\eta, \zeta \in \mathfrak{g}.
\end{equation}
For example, if $G $ is compact such an inner product always exists. If $G $ is
reductive one can always find such a nondegenerate $\gamma$ but it may be
indefinite.

Given a Riemannian metric $g$ on $M$ and an
$\operatorname{Ad}$-invariant inner product
$\gamma$ on $\mathfrak{g}$ we can
define a Riemannian metric $g\gamma$ on the vector bundles
$\Lambda^k(M,\operatorname{Ad}P)\rightarrow M$ of $\operatorname{Ad}P$-valued
exterior $k $-forms on $M $. Indeed, the inner product
$\gamma$ induces a Riemannian metric on the vector bundle
$\operatorname{Ad}P\rightarrow M$ whose value on 
$[p,\xi]_G,[p,\eta]_G\in(\operatorname{Ad}P)_x$, $x : = \pi(p)$, $p \in P $, is
given by
\[
\gamma _x\left( [p,\xi]_G,[p,\eta]_G\right) := \gamma(\xi,\eta).
\]
Denote, by abuse of notation, by the same letter $ \gamma $ the smooth vector
bundle metric on $ \operatorname{ Ad}P $ defined by $ \gamma
|_{(\operatorname{Ad}P)_x}: = \gamma _x $. Let $\overline{g}$ denote the
Riemannian metric induced by $ g $ on the vector bundles $\Lambda^kM\rightarrow
M$ of exterior $k$-forms on $M$. The Riemannian metric $g\gamma$ on the vector
bundle
$\Lambda^k(M,\operatorname{Ad}P)\rightarrow M$ is then constructed in the
following manner. If $\alpha_x, \beta_x\in\Lambda^k(M,\operatorname{Ad}P)_x$,
write
$\alpha_x=\alpha^af_a$ and $\beta_x=\beta^af_a$, where $\{f_a\}$ is a
basis of the fiber $(\operatorname{Ad}P)_x$, and
$\alpha^a,\beta^a\in(\Lambda^k M)_x$. Then define
\[
(g\gamma)_x(\alpha_x,\beta_x):=\gamma_{ab}\,\overline{g}(\alpha^a,\beta^b),
\]
where $\gamma_{ab}:=\gamma_x(f_a,f_b)$. It is easy to verify that this
construction is independent on the choice of the basis in each fiber
$(\operatorname{Ad}P)_x$.
\medskip

Let $M$ be a compact oriented boundaryless manifold. If
$\alpha\in\Omega^k(M,\operatorname{Ad}P)$ and
$\beta\in\Omega^{k+1}(M,\operatorname{Ad}P)$ we have (see, e.g.,  Theorem 4.2.9 in \cite{Bleecker1981}):
\begin{equation}\label{adjoint}
\int_M(g\gamma)(\textbf{d}^\mathcal{A}\alpha,\beta)\mu=\int_M(g\gamma)(\alpha,\delta^\mathcal{A}\beta)\mu,
\end{equation}
where $\mu$ denotes the volume form associated to the Riemannian metric $g$.

Given a connection $\mathcal{A}$, a Riemannian metric $g$ on $M$, and an
$\operatorname{Ad}$-invariant inner product $\gamma$ on $\mathfrak{g}$, we can
define the \textbf{Kaluza-Klein metric} $K_\mathcal{A}$ on
$TP$ by
\begin{equation}
\label{KK_metric}
K_\mathcal{A}(u_p,v_p):=g(T_p\pi(u_p),T_p\pi(v_p))+\gamma(\mathcal{A}(u_p),\mathcal{A}(v_p)).
\end{equation}
The Kaluza-Klein metric is $G $-invariant, that is,
$\Phi_g^*K_\mathcal{A}=K_\mathcal{A}$ for any $g \in G $.

\subsection{Expressions in a local trivialization} Consider a local
trivialization $P_U:=\pi^{-1}(U)\rightarrow U\times
G,\;\;p\mapsto (x,g)$. This induces a local
trivialization of the vector bundle $\operatorname{Ad}P\rightarrow
M$, given by
\begin{equation}\label{local coordinates}
\operatorname{Ad}P\supset P_U\times_G\mathfrak{g}\rightarrow
U\times\mathfrak{g},\;\;[(x,g),\xi]_G\mapsto (x,\operatorname{Ad}_g\xi).
\end{equation}
It is useful to note that for $\omega\in\overline{\Omega^k}(P,\mathfrak{g})$,
we
can locally write
\[
\omega(x,g)((v_1,a_1),...,(v_k,a_k))=\operatorname{Ad}_{g^{-1}}(\overline{\omega}(x)(v_1,...,v_k)),
\]
where
$\overline{\omega}(x)(v_1,...,v_k):=\omega(x,e)((v_1,0),...,(v_k,0))
\in \mathfrak{g}$,
$v_i \in T_xU $, $a_i \in T_g G $, $i = 1, \dots ,k$. Moreover, in the
local trivialization \eqref{local coordinates}, the $\operatorname{Ad}P$-valued
$k$-form $\widetilde{\omega}\in\Omega^k(M,\operatorname{Ad}P)$
defined in \eqref{tilde} is given by
\[
\tilde{ \omega}(x)(v_1,...,v_k) =
(x,\overline{\omega}(x)(v_1,...,v_k)), \quad x \in U , \quad v_i
\in T_xU.
\]
Recall also that, in a local trivialization, a connection
$\mathcal{A}$ can be written as
\begin{equation}\label{localconnection}
\mathcal{A}(x,g)(v_x,\xi_g)=\operatorname{Ad}_{g^{-1}}(\overline{\mathcal{A}}(x)(v_x)+TR_{g^{-1}}(\xi_g)),
\end{equation}
where $v_x \in T_xU$, $\xi_g \in T_g G $, and
$\overline{\mathcal{A}}$ is a
$1$-form on
$U\subset M$ with values in $\mathfrak{g}$. Locally, for
$f\in\mathcal{F}_G(P,\mathfrak{g})$ and the curvature
$\mathcal{B}$, we can write
\begin{align*}
\overline{\textbf{d}^\mathcal{A}f}(v)&=\textbf{d}\overline{f}(v)+[\overline{\mathcal{A}}(v),\overline{f}],\\
\overline{\mathcal{B}}(v,w)&=\textbf{d}\overline{\mathcal{A}}(v,w)+[\overline{\mathcal{A}}(v),\overline{\mathcal{A}}(w)].
\end{align*}
If the principal bundle is trivial, the previous formulas hold
globally and the adjoint bundle is also trivial
$\operatorname{Ad}P\cong M\times \mathfrak{g}$, so we have
$\Omega^k(M,\operatorname{Ad}P)=\Omega^k(M,\mathfrak{g})$ and for
$\omega\in\overline{\Omega^k}(P,\mathfrak{g})$ we get
$\widetilde{\omega}=\overline{\omega}$.

\subsection{Automorphisms and gauge transformations} 
We say that a diffeomorphism $\varphi$ of $P$ is an
\textbf{automorphism} if it is equivariant, that  is, 
$\Phi_g\circ\varphi=\varphi\circ\Phi_g$,
for all $g\in G$. The Fr\'echet Lie group of all
automorphisms is denoted by
$\mathcal{A}ut(P)$. See \cite{KrMi1997} for an account of Fr\'echet Lie groups
in the framework of manifold of maps from the point of view of the ``convenient
calculus". An automorphism $\varphi$ of $P$ induces a
unique diffeomorphism
$\overline{\varphi}$ of $M$ defined by the condition
$\pi\circ\varphi=\overline{\varphi}\circ\pi$. The Lie algebra
$\mathfrak{aut}(P)$ consists of $G$-invariant vector fields on
$P$. Its (left) Lie bracket is denoted by $[U,V]_L$ and is the
negative of the usual Jacobi-Lie bracket $[U,V]_{JL}$. For
$U\in\mathfrak{aut}(P)$ we denote by $[U]\in\mathfrak{X}(M)$ the
unique vector field on $M$ defined by the condition
\[
T\pi\circ U=[U]\circ\pi.
\]
The subgroup $\mathcal{A}ut_\mu(P)$ consists, by definition, of automorphisms
$\varphi$ of $P $ such that the induced diffeomorphism  $\overline{\varphi}$
preserves the volume form $\mu$ on $M$.  For
$U\in\mathfrak{aut}_\mu(P)$ we have $[U]\in\mathfrak{X}_{\rm div}(M)$, the space
of
all divergence free vector fields on $M$. 

The normal subgroup $\mathcal{G}au(P)$ of \textbf{gauge transformations}
contains, by definition, all automorphisms $\varphi$ on $P $ with
$\overline{\varphi}=id_M$. Note that we can identify the gauge group
$\mathcal{G}au(P)$ with the group
\[
\mathcal{F}_G(P,G):=\{\tau\in\mathcal{F}(P,G)\mid
\tau\circ\Phi_g=\operatorname{AD}_{g^{-1}}\circ\,\tau\},\;\text{where}\;\operatorname{AD}_g(h):=ghg^{-1}.
\]
The identification is given by the group isomorphism
$\widehat\,:\mathcal{G}au(P)\rightarrow\mathcal{F}_G(P,G)$, which associates to
$\varphi\in\mathcal{G}au(P)$, the map $\widehat\varphi\in\mathcal{F}_G(P,G)$
defined by the condition
\[
\varphi(p)=\Phi_{\widehat\varphi(p)}(p).
\]

The Lie algebra $\mathfrak{gau}(P)$ consists of $G$-invariant vertical vector
fields on $P$. Therefore when $U\in\mathfrak{gau}(P)$ we have $[U]=0$. Note the
identifications
\[
\mathfrak{gau}(P)\cong\mathcal{F}_G(P,\mathfrak{g})\cong\Gamma(\operatorname{Ad}P).
\]
Indeed, to $f\in\mathcal{F}_G(P,\mathfrak{g})$ we can associate the
$G$-invariant vertical vector field $\sigma(f) \in \mathfrak{gau}(P)$ given by
\begin{equation}\label{sigma}
\sigma(f)(p):=f(p)_P(p).
\end{equation}
The second isomorphism is given by the map \eqref{tilde}. A direct computation
shows  that $\sigma: \mathcal{F}_G(P, \mathfrak{g}) \rightarrow
\mathfrak{gau}(P)$ is a Lie algebra isomorphism, that is,
\[
\sigma([f,g])=[\sigma(f),\sigma(g)]_L.
\]

The transformation law of a connection $\mathcal{A}$
under $\varphi \in \mathcal{G}au(P)$ is given by
\begin{equation}
\label{conn_gau_transf}
\varphi^* \mathcal{A} = \operatorname{Ad}_{\widehat{ \varphi} ^{-1}} \circ
\mathcal{A} + TL_{\widehat{ \varphi} ^{-1}} \circ T \widehat{ \varphi} .
\end{equation}

If the principal bundle  $P \rightarrow M $  is trivial the automorphism group
is the semidirect product of $ \mathcal{D}(M) $ with $\mathcal{F}(M,G)$. To see
this, note first that each
$\varphi\in\mathcal{A}ut(P)$ is in this case of the form 
\[
\varphi(x,g)=(\overline{\varphi}(x),\overline{\overline{\varphi}}(x)g),
\]
where $\overline{\varphi} \in \mathcal{D}(M)$, the diffeomorphism group of $M
$,
and $\overline{\overline{\varphi}}\in\mathcal{F}(M,G)$, the smooth $G $-valued
functions on $M $. Thus the map $\varphi \in \mathcal{A}ut(P) 
\mapsto (\overline{\varphi}, \overline{\overline{\varphi}}) \in \mathcal{D}(M)
\times \mathcal{F}(M, G)$ is bijective. Second, the pair 
$\left(\overline{\varphi_1 \circ \varphi_2}, \overline{\overline{\varphi_1
\circ
\varphi_2}} \right)$ corresponding to the product $\varphi_1\circ\varphi_2$ is
uniquely determined by the right hand side of the identity
\[
\left(\varphi_1 \circ \varphi_2 \right)(x, g) = 
\left((\overline{\varphi_1}\circ\overline{\varphi_2})(x),
(\overline{\overline{\varphi_1}}\circ\overline{\varphi_2})(x)\overline{\overline{\varphi_2}}(x)
g \right).
\]
This shows that  the map $\varphi \in \mathcal{A}ut(P) \mapsto (
\overline{\varphi}, \overline{\overline{\varphi}}) \in \mathcal{D}(M)
\,\circledS\, \mathcal{F}(M, G)$ is a group isomorphism, where the semidirect
product is defined by the right action of $\mathcal{D}(M)$ by group
automorphisms on $\mathcal{F}(M, G)$ given by $( \chi, \lambda) \in
\mathcal{F}(M, G) \times \mathcal{D}(M) \mapsto \chi \circ \lambda \in
\mathcal{F}(M, G)$. In particular, if $\psi \in \mathcal{G}au(P)$, then
$\overline{ \psi} = id_M $ and we have $\psi(x, g) = ( x,
\overline{\overline{\psi}}(x)g)$, which shows that the map $\psi \in
\mathcal{G}au(P) \mapsto \overline{\overline{\psi}} \in \mathcal{F}(M, G)$ is a
group isomorphism. 

The same considerations hold for the volume preserving case. We have shown
hence
that \textit{if the principal $G $-bundle $\pi: P \rightarrow M$ is trivial,
then we have the group isomorphisms} 
\[
\mathcal{A}ut(P)\simeq\mathcal{D}(M)\,\circledS\,\mathcal{G}au(P)\quad\text{and}\quad\mathcal{A}ut_\mu(P)\simeq\mathcal{D}_\mu(M)\,\circledS\,\mathcal{G}au(P)
\]
\textit{and the corresponding Lie algebra isomorphisms}
\[
\mathfrak{aut}(P) \simeq
\mathfrak{X}(M)\,\circledS\,\mathcal{F}(M,\mathfrak{g})\quad\text{and}\quad\mathfrak{aut}_\mu(P)
\simeq \mathfrak{X}_{\rm div}(M)\,\circledS\,\mathcal{F}(M,\mathfrak{g}).
\]
Using the general formula for the Lie bracket associated the Lie algebra of a
semidirect product of two groups (see formula (6.4.2) in \cite{MaMiOrPeRa2007},
for example), we find that the (left) Lie bracket on $\mathfrak{aut}(P)$ and
$\mathfrak{aut}_\mu(P)$ is 
\begin{equation}\label{Liebracket}
[(v,\theta),(w,\omega)]_L=\left([v,w]_L,\textbf{d}\theta(w)-\textbf{d}\omega(v)+[\theta,\omega]\right).
\end{equation}

If the principal bundle $P \rightarrow M $ is not trivial, the situation is
more
involved. First, the sequence
\[
0 \longrightarrow \mathfrak{gau}(P) \longrightarrow \mathfrak{aut}(P)
\longrightarrow \mathfrak{X}(M) \longrightarrow 0
\]
is exact. The second arrow is the inclusion and the third is the Lie algebra
homomorphism given by $U \in \mathfrak{aut}(M) \mapsto [U] \in \mathfrak{X}(M)$
which is surjective because any $X \in \mathfrak{X}(M)$ is covered by its
horizontal lift relative to some connection. Note, however, that the horizontal
lift of vector fields relative to a connection is, in general, not a Lie
algebra
homomorphism since the bracket of two horizontally lifted vector  fields has a
vertical part. This is an indication that if $P \rightarrow M $ is nontrivial, then
$\mathfrak{aut}(P)$ is, in general, not the semidirect product of
$\mathfrak{X}(M) $ with $\mathfrak{gau}(P)$.

Second, at group level, the
map
$\mathcal{A}ut(P) \rightarrow \mathcal{D}(M)$ is not surjective, in general.
For
example, let $P = S^3 \subset  \mathbb{R}^4$ be the unit sphere, thought of as
the unit quaternions, and let $S^1 : = \{x +y
\mathbf{k} \mid x, y \in \mathbb{R}, x^2 + y^2 = 1\}$ act on $P $ by $q \mapsto
q(x + y \mathbf{k})$. The Hopf fibration map $\pi: q \in S ^3 \mapsto q
\mathbf{k} \overline{q} \in S^2$ defines a principal $S^1$-bundle over $M : =
S^2$. We shall prove that in this case the map $\mathcal{A}ut(P) \rightarrow
\mathcal{D}(M)$ is not surjective.\footnote{We thank Marco Castrill\'on-L\'opez
for this example.} Let $\eta \in \mathcal{D}(S^2)$ be the antipodal map whose
degree is -1 and is hence not homotopic to the identity. If there were some
$\varphi \in \mathcal{A}ut(S^3)$ descending to $\eta$, then $\varphi$ would not
have any fixed points and hence its degree would be one. By the Hopf Degree
Theorem $\varphi$ would then be homotopic to the identity which would imply
that
$\eta$ was homotopic to the identity, a contradiction.

\subsection{Duality} In this paper we will identify the cotangent space
$T_\varphi^*\mathcal{A}ut(P)$ with the space of $G$-invariant $1$-forms on $P$
along $\varphi\in\mathcal{A}ut(P)$. The duality pairing is 
\[
\left\langle \mathbf{M}_\varphi, U _\varphi \right\rangle : = \int_M
\mathbf{M}_\varphi(U _\varphi) \mu,
\]
where $\mathbf{M}_\varphi \in T ^\ast_\varphi \mathcal{A}ut (P)$ and $U
_\varphi
\in T _\varphi \mathcal{A}ut (P)$. Note that  in this formula we used the fact
that $ \mathbf{M}_\varphi(U _\varphi)  $ is a smooth function on $P $ that does
not depend on the fiber variables and hence  induces a unique smooth function
on
$M $ which is then integrated using the volume form $\mu$ on $M $. In
particular
we have $\mathfrak{aut}(P)^*=\Omega^1_G(P)$, the space of
right-invariant $1$-forms on $P$.

We identify the cotangent space $T^*_\varphi\mathcal{G}au(P)$ with the tangent
space $T_{\varphi} \mathcal{G}au(P)$ via the duality
\begin{equation}\label{duality_for_Gau}
\left\langle
U_\varphi,V_\varphi\right\rangle:=\int_M\gamma\left(\widetilde{\mathcal{A}(U_\varphi)},\widetilde{\mathcal{A}(V_\varphi)}\right)\mu,
\end{equation}
for any principal connection $\mathcal{A}$ on $P$. Note that, since $U_\varphi$
and $V_\varphi$ are vertical, the pairing \eqref{duality_for_Gau} does not
depend on $\mathcal{A}$ since 
$\mathcal{A}(U_\varphi)=\sigma^{-1}(U_\varphi\circ\varphi^{-1})\circ\varphi$
for
any connection $\mathcal{A}$.

\section{Equations for the fields}\label{Fields}

In this section we give the Lagrangian and Hamiltonian formulations for the
Yang-Mills fields in the vacuum. We will see that it is not possible to pass
from one to the other by a simple Legendre transformation.

\subsection{Lagrangian formulation of the Maxwell equations}
\label{Lagrangian formulation of the Maxwell equations}

On the Lagrangian side, the variables are the magnetic potential
$A\in\Omega^1(M)$ and the electric potential
$A_0\in\mathcal{F}(M)$, where $M $ is a three dimensional
compact manifold without boundary. The Lagrangian is defined
on the tangent bundle
$T(\mathcal{F}(M)\times\Omega^1(M))$ and is given
by
\[
L(A_0,\dot{A}_0,A,\dot{A})=\frac{1}{2}\int_M\|E\|^2\mu-\frac{1}{2}\int_M\|B\|^2\mu,
\]
where $E:=-\dot{A}+\textbf{d}A_0$, $B:=\textbf{d}A$, and
$\| \cdot \|$ is the norm associated to the Riemannian metric
induced by $g$ on the vector bundle $\Lambda^kM\rightarrow
M$, for $k = 1,2$. The Euler-Lagrange equations associated to $L$ are 
\begin{equation}
\label{EL_Maxwell}
\delta E=0\qquad \text{and}\qquad \displaystyle\frac{\partial E}{\partial
t}=\delta B.
\end{equation}
The relations $E=-\dot{A}+\textbf{d}A_0$ and $B=\textbf{d}A$ give
\begin{equation}
\label{rest_Maxwell}
\displaystyle\frac{\partial B}{\partial
t}=-\textbf{d}E\;\;\;\text{and}\;\;\;\textbf{d}B=0.
\end{equation}
Using the vector field variables $\textbf{E}:=E^\sharp$ and
$\textbf{B}:=(\star B)^\sharp$, where $\star: \Omega^k(M) \rightarrow
\Omega^{3-k}(M) $, is the Hodge-star operator associated to the Riemannian
metric $g $ on $M $,   we obtain the Maxwell
equations in the vacuum
\begin{equation}\label{Maxwell}
\left\lbrace
\begin{array}{ll}
\vspace{0.2cm}\displaystyle\frac{\partial \textbf{E}}{\partial
t}=\operatorname{curl}\textbf{B},&\operatorname{div}\textbf{E}=0,\\
\displaystyle\frac{\partial \textbf{B}}{\partial
t}=-\operatorname{curl}\textbf{E},&\operatorname{div}\textbf{B}=0,
\end{array}
\right.
\end{equation}
where $\operatorname{curl}: \mathfrak{X}(M) \rightarrow \mathfrak{X}(M)$ is the
operator $\operatorname{curl}(\mathbf{X}): = [\star ( \mathbf{d}
\mathbf{X}^\flat)]^\sharp$, for any $\mathbf{X} \in \mathfrak{X}(M)$.
\medskip

Let us recall the classical argument  that we can choose $A_0 = 0$. Assume that
$A'_0$ and $A'$ satisfy Maxwell's equations \eqref{EL_Maxwell} and
\eqref{rest_Maxwell}. We search a function $\varphi \in \mathcal{F}(M)$ such
that $A: = A' + \mathbf{d}\varphi$ leaves the equations \eqref{EL_Maxwell} and
\eqref{rest_Maxwell} unchanged and  $A_0 = 0$. Since these equations are second
order we have $\dot{A'}_0 = \partial A'_0/  \partial t$ and $\dot{A'} =
\partial
A'/  \partial t$. Let $E':=-\dot{A}'+\textbf{d}A'_0$, $B':=\textbf{d}A'$. The
requirement is that $E' = E $ and $B' = B$. Therefore,
\[
-\dot{A'}+\textbf{d}A'_0 =
E' = E = -\dot{A}+\textbf{d}A_0
= - \dot{A'} - \mathbf{d}\dot{ \varphi} + \mathbf{d}A_0
\]
which is equivalent to $\mathbf{d}\dot{ \varphi} = \mathbf{d}(A_0 - A'_0)$ and
hence it is sufficient to choose $A_0: = A'_0 + \dot{\varphi}$.
This shows that one can choose $A_0 = 0$ provided $\dot{ \varphi} = - A'_0$.
Note that the equations \eqref{EL_Maxwell} and
\eqref{rest_Maxwell} are unchanged under this transformation, as required.
\medskip

We now recall the four dimensional formulation of the Maxwell equations.
Consider the Lorentzian manifold $(X,\gamma)$ given by $X=M\times\mathbb{R}$
and
$\gamma:=\tau_1^*g-\tau_2^*dt^2$, where $\tau_1 : X\rightarrow M$ and $\tau_2 :
X\rightarrow \mathbb{R}$ are the natural projections and $g$ is a Riemannian
metric on $M$. Consider the $1$-form $G$ on $X$ defined by
$G:=\tau_1^*A_t+\tau_1^*A_0\wedge \tau_2^*dt$. We have
\[
\textbf{d} G=\tau_1^*\textbf{d}A_t-\tau_1^*\dot{A}_t\wedge
\tau_2^*dt+\textbf{d}A_0\wedge \tau_2^*dt=\tau_1^*B+\tau_1^*E\wedge 
\tau_2^*dt=:F,
\]
and the Maxwell equations \eqref{Maxwell} can be simply written as (see e.g.
\cite{MiThWh1973}, \S22.4)
\[
\textbf{d}F=0\;\;\;\text{and}\;\;\;\delta F=0.
\]
In a general slicing of space-time, not just $M \times \mathbb{R}$, the
derivation of these equations and much more information can be found, for
example, in \cite{GoIsMa1997, GoIsMa1999}.

The Legendre transformation associated to the Maxwell Lagrangian $L$ is not
bijective. Thus, it is not possible to pass in the usual way from the Lagrangian
to the Hamiltonian formulation by the Legendre transformation.
This degeneracy is typical of relativistic field theories and is resolved by the
Dirac theory of constraints; see, for example, \cite{GoIsMa1997, GoIsMa1999} and
references therein. In the next subsection we
directly generate the Hamiltonian formulation for the Maxwell equations.

\subsection{Hamiltonian formulation of the Maxwell equations}

On the Hamiltonian side (see \cite{MaWeRaScSp1983}), the configuration space
variable is the magnetic potential $A\in\Omega^1(M)$. The Hamiltonian is
defined
on the cotangent bundle $T^*\Omega^1(M) \simeq \Omega^1(M) \times \Omega^1(M)$,
where the cotangent space at any point $A $ is identified with $\Omega^1(M)$
using the natural $L^2$-pairing,  and is given by
\[
H(A,Y)=\frac{1}{2}\int_M\|E\|^2\mu+\frac{1}{2}\int_M\|B\|^2\mu
\]
for $E:=-Y$ and $B:=\textbf{d}A$. Hamilton's equations are
\[
\frac{\partial B}{\partial t}=-\textbf{d}E\;\;\;\text{and}\;\;\;\frac{\partial
E}{\partial t}=\delta B,
\]
and the relation $B=\textbf{d}A$ gives
\[
\textbf{d}B=0.
\]
To obtain the last equation $\delta E=0$ we use the invariance of the
Hamiltonian under gauge transformations. The action of the gauge group
$\mathcal{F}(M)$ on $\Omega^1(M)$ is given by
\begin{equation}
\label{gauge_group_Maxwell}
\mathcal{F}(M)\times\Omega^1(M)\rightarrow
\Omega^1(M),\qquad (\varphi,A)\mapsto A+\textbf{d}\varphi,
\end{equation}
and is Hamiltonian. The associated momentum map is
\[
\textbf{J} :
T^*\Omega^1(M)\rightarrow\mathcal{F}(M)^* \simeq \mathcal{F}(M) ,\qquad
\textbf{J}(A,Y)=\delta Y,
\]
where $\mathcal{F}(M)^*$ is identified with $\mathcal{F}(M)$ using the natural
$L^2$-pairing. So the condition $\textbf{J}(A,Y)=0$ gives the fourth Maxwell
equation $\delta E=0$.

Note that in the Hamiltonian formulation we have used only the configuration
variable $A $, whereas in the Lagrangian formulation the configuration space
consisted of pairs $(A_0, A)$. As we have seen, the variable $A_0$ can be set
equal to zero  without any effect on Maxwell's equations. Note also that the
Euler-Lagrange equation $\delta E = 0$ was obtained from the variation of the
Lagrangian relative to $A_0$, whereas in the Hamiltonian set-up this equation
appears as a conservation law for the gauge group action
\eqref{gauge_group_Maxwell}.

\subsection{Generalization to any principal bundle}

We now generalize the previous formulations to the case of a $G$-principal
bundle
$P\rightarrow M$ over an arbitrary compact boundaryless
manifold $M$. We will show that if  $M $ is three
dimensional, $G=S^1$, and the bundle is trivial, then we recover the Maxwell
equations.

\medskip \noindent{\bf Lagrangian formulation.} The
Lagrangian  $L: T(\mathcal{F}_G(P,\mathfrak{g})\times\mathcal{C}onn(P))
\rightarrow \mathbb{R}$
is defined by
\begin{equation}\label{YMLagrangian}
L(\mathcal{A}_0,\dot{\mathcal{A}}_0,\mathcal{A},\dot{\mathcal{A}})
=\frac{1}{2}\int_M\|E\|^2\mu-\frac{1}{2}\int_M\|B\|^2\mu,
\end{equation}
where:
\begin{enumerate}
\item[(1)] $E:=\widetilde{\mathcal{E}}\in\Omega^1(M,\operatorname{Ad}P)$ is the
$\operatorname{Ad}P$-valued $1$-form associated, through the map
$\eqref{tilde}$, to the
``electric part"
$\mathcal{E}\in\overline{\Omega^1}(P,\mathfrak{g})$ of the
Yang-Mills field, given by
\[
\mathcal{E}:=-\dot{\mathcal{A}}+\textbf{d}^\mathcal{A}\mathcal{A}_0\in\overline{\Omega^1}(P,\mathfrak{g});
\]
\item[(2)] $B:=\widetilde{\mathcal{B}}\in\Omega^2(M,\operatorname{Ad}P)$ is the
$\operatorname{Ad}P$-valued $2$-form associated, through the map
$\eqref{tilde}$, to the
``magnetic part" $\mathcal{B}\in\overline{\Omega^2}(P,\mathfrak{g})$ of
the
Yang-Mills field, given by the curvature
\[
\mathcal{B}:=\textbf{d}^\mathcal{A}\mathcal{A}\in\overline{\Omega^2}(P,\mathfrak{g});
\]
\item[(3)] $\|\cdot \|$ is the norm associated to the metric
$g\gamma$ on the vector bundles
$\Lambda^k(M,\operatorname{Ad}P)\rightarrow M$, for $k = 1,2$.
\end{enumerate}
\medskip

The Euler-Lagrange equations associated to $L$ are 
\[
\delta^\mathcal{A}
\mathcal{E}=0\;\;\;\text{and}\;\;\;\displaystyle\frac{\partial
\mathcal{E}}{\partial t}+[\mathcal{A}_0,\mathcal{E}]=\delta^\mathcal{A}
\mathcal{B}.
\]
Indeed, using the $L^2$ pairing
\begin{equation}\label{L^2pairing}
\langle
\alpha,\beta\rangle=\int_M(g\gamma)(\alpha,\beta)\mu,\qquad
\alpha,\beta\in\Omega^k(M,\operatorname{Ad}P),
\end{equation}
we can identify the cotangent bundles of $\mathcal{F}_G(P, \mathfrak{g})$ and
$\mathcal{C}onn(P)$ with their tangent bundles. Using formulas
\eqref{Ad-invariance}, \eqref{adjoint}, and the identity
\begin{equation}
\label{der_connection}
\left.\frac{d}{dt}\right|_{t=0}\textbf{d}^{\mathcal{A}+t\mathcal{C}}(\mathcal{A}+t\mathcal{C})=\textbf{d}^\mathcal{A}\mathcal{C},
\end{equation}
where $\mathcal{A}\in\mathcal{C}onn(P)$, $\mathcal{C}\in
T_\mathcal{A}\mathcal{C}onn(P)=\overline{\Omega^1}(P,\mathfrak{g})$,
we get
\[
\frac{\partial
L}{\partial\mathcal{A}_0}=\delta^\mathcal{A}\mathcal{E},\quad \frac{\partial
L}{\partial\dot{\mathcal{A}}_0}=0,\quad \frac{\partial
L}{\partial\mathcal{A}}=-\delta^\mathcal{A}\mathcal{B}+[\mathcal{A}_0,\mathcal{E}],\quad
\frac{\partial
L}{\partial\dot{\mathcal{A}}}=-\mathcal{E}.
\]
Thus, the Euler-Lagrange equations
\[
\frac{\partial}{\partial t}\frac{\partial
L}{\partial\dot{\mathcal{A}}_0}-\frac{\partial
L}{\partial\mathcal{A}_0}=0\quad\text{and}\quad\frac{\partial}{\partial
t}\frac{\partial
L}{\partial\dot{\mathcal{A}}}-\frac{\partial
L}{\partial\mathcal{A}}=0
\]
become
\[
\delta^\mathcal{A}
\mathcal{E}=0\;\;\;\text{and}\;\;\;\displaystyle\frac{\partial
\mathcal{E}}{\partial t}+[\mathcal{A}_0,\mathcal{E}]=\delta^\mathcal{A}
\mathcal{B},
\]
as stated above.
\medskip

The relations
$\mathcal{E}:=-\dot{\mathcal{A}}+\textbf{d}^\mathcal{A}\mathcal{A}_0$ and
$\mathcal{B}:=\textbf{d}^\mathcal{A}\mathcal{A}$ give the equations
\[
\displaystyle\frac{\partial \mathcal{B}}{\partial
t}+[\mathcal{A}_0,\mathcal{B}]=-\textbf{d}^\mathcal{A}\mathcal{E}\qquad
\text{and}\qquad \textbf{d}^\mathcal{A}\mathcal{B}=0.
\]
Indeed, for the first equality we have, using \eqref{der_connection} and Lemma
\ref{d^Ad^A},
\[
\dot{\mathcal{B}}=\textbf{d}^\mathcal{A}\dot{\mathcal{A}}=-\textbf{d}^\mathcal{A}\mathcal{E}+\textbf{d}^\mathcal{A}\textbf{d}^\mathcal{A}\mathcal{A}_0=-\textbf{d}^\mathcal{A}\mathcal{E}+[\mathcal{B},\mathcal{A}_0].
\]
The second equality is the Bianchi identity. Summarizing, we get the system
\begin{equation}\label{YM}
\left\lbrace
\begin{array}{ll}
\vspace{0.2cm}\displaystyle\frac{\partial \mathcal{E}}{\partial
t}+[\mathcal{A}_0,\mathcal{E}]=\delta^\mathcal{A}
\mathcal{B},&\delta^\mathcal{A}\mathcal{E}=0,\\
\displaystyle\frac{\partial \mathcal{B}}{\partial
t}+[\mathcal{A}_0,\mathcal{B}]=-\textbf{d}^\mathcal{A}\mathcal{E},&\textbf{d}^\mathcal{A}\mathcal{B}=0.
\end{array}
\right.
\end{equation}

To recover Maxwell's equations we take a trivial $S^1$-principal bundle $P = M
\times S ^1$. Then $\operatorname{Ad}P=M\times\mathbb{R}$ and
$\Omega^k(M,\operatorname{Ad}P)=\Omega^k(M)$. Since the structure group $S^1$
of
the principal bundle $P $ is
Abelian, the covariant differential does not depend on the connection, that is,
$\textbf{d}^\mathcal{A}=\textbf{d}$. We obtain the following identifications.
\begin{itemize}
\item [(1)] From the equality
$\mathcal{E}=-\dot{\mathcal{A}}+\textbf{d}^\mathcal{A}\mathcal{A}_0$, we obtain
that the electric field $E:=\widetilde{\mathcal{E}}\in\Omega^1(M)$ is given by
\[
E=-\dot{A}+\textbf{d}A_0,
\]
where $A_0:=\widetilde{\mathcal{A}_0}\in\mathcal{F}(M)$ and
$\dot{A}:=\widetilde{\dot{\mathcal{A}}_{\;}}\in\Omega^1(M)$.

\item[(2)] From the equality $\mathcal{B}=\textbf{d}^\mathcal{A}\mathcal{A}$,
we
obtain that the magnetic field
$B:=\widetilde{\mathcal{B}}\in\Omega^2(M)$ is given by
\[
B=\textbf{d}A,
\]
where $A\in\Omega^1(M)$ is given by $A:=\overline{\mathcal{A}}$ (see
equation \eqref{localconnection}).
\end{itemize}

Returning to the general case, let us show, as in the case of Maxwell's
equations, that we can choose
$\mathcal{A}_0 = 0$. Assume that $\mathcal{A}'_0$ and $\mathcal{A}'$ satisfy
equations \eqref{YM}. We search a  $\varphi \in \mathcal{G}au(P)$ such that
$\mathcal{A}: = \varphi^\ast\mathcal{A}'$ leaves the equations \eqref{YM}
unchanged and  $\mathcal{A}_0 = 0$. Since these equations are second order we
have $\dot{\mathcal{A}}'_0 = \partial \mathcal{A}'_0/  \partial t$ and
$\dot{\mathcal{A}}' = \partial \mathcal{A}'/  \partial t$. Let
$\mathcal{E}':=-\dot{\mathcal{A}}'+\textbf{d}^{ \mathcal{A}'}\mathcal{A}'_0$,
$\mathcal{B}':=\textbf{d}^{ \mathcal{A}}\mathcal{A}'$. Since
$\mathcal{B}=\varphi^\ast\mathcal{B}'$, the requirement is that
$\mathcal{E}=\varphi^*\mathcal{E}'$. Therefore, emphasizing the
time-dependence,
we have the equivalences
\begin{align}\label{equivalences}
&-\dot{\mathcal{A}}_t+\textbf{d}^{
\mathcal{A}_t}\mathcal{A}_{0t}=\varphi_t^*\left(-\dot{\mathcal{A}}_t'+\textbf{d}^{
\mathcal{A}'_t}\mathcal{A}'_{0t}\right) \\
\Longleftrightarrow \quad & -\frac{\partial}{\partial
t}(\varphi_t^*\mathcal{A}'_t)+\textbf{d}^{
\mathcal{A}_t}\mathcal{A}_{0t}=-\varphi_t^\ast\frac{\partial}{\partial
t}\mathcal{A}_t'+\textbf{d}^{\mathcal{A}_t}\varphi_t^*\mathcal{A}_{0t}'\nonumber\\
\Longleftrightarrow \quad & \textbf{d}^{
\mathcal{A}_t}\mathcal{A}_{0t}=\left.\frac{\partial}{\partial
t}\right|_{t=s}\varphi_s^\ast\mathcal{A}'_t+\textbf{d}^{\mathcal{A}_t}\varphi_t^*\mathcal{A}_{0t}'. \nonumber
\end{align}
Taking the time derivative of \eqref{conn_gau_transf} we get
\begin{equation}
\label{der_gau_connection}
\left.\frac{d}{dt}\right|_{t=0}\psi_t^*\mathcal{A}=\textbf{d}^\mathcal{A}\left(\sigma^{-1}(\dot{\psi_0})\right),
\end{equation}
for any smooth curve $\psi_t\in\mathcal{G}au(P)$ such that $\psi_0=id$.
Therefore we conclude that
\begin{align*} 
\left.\frac{\partial}{\partial
s}\right|_{s=t}\varphi_s^*\mathcal{A}'_t&=\left.\frac{\partial}{\partial
s}\right|_{s=t}
\varphi_t^*(\varphi_s\circ\varphi_t^{-1})^*\mathcal{A}'_t=\varphi_t^*\textbf{d}^{\mathcal{A}'_t}\left(\sigma^{-1}(\dot{\varphi}_t\circ\varphi_t^{-1})\right) \\
&=\textbf{d}^{\mathcal{A}_t}\varphi_t^*\left(\sigma^{-1}(\dot{\varphi}_t\circ\varphi_t^{-1})\right).
\end{align*} 
So \eqref{equivalences} is equivalent to $\textbf{d}^{
\mathcal{A}_t}\mathcal{A}_{0t}=\textbf{d}^{\mathcal{A}_t}\varphi_t^*\left(\sigma^{-1}(\dot{\varphi}_t\circ\varphi_t^{-1})+\mathcal{A}_{0t}'\right)$
and hence it is sufficient to choose
$\mathcal{A}_{0t}:=\varphi_t^*\left(\sigma^{-1}(\dot{\varphi}_t\circ\varphi_t^{-1})+\mathcal{A}_{0t}'\right)$
in order to get $\mathcal{E}=\varphi^*\mathcal{E}'$.
Thus one can choose $\mathcal{A}_{0t}=0$ provided that
$\dot{\varphi}_t\circ\varphi_t^{-1}=-\sigma(\mathcal{A}_{0t}')$. A direct
computation shows that the equations \eqref{YM} are unchanged under this
transformation, as required.

\medskip

Let $(X= M \times \mathbb{R},\gamma)$ be the Lorentzian manifold given in
\S\ref{Lagrangian formulation of the Maxwell equations}. Let
$\overline{P}:=P\times \mathbb{R}$ and define the free $G $-action
$\overline{\Phi}_g(p,t):=(\Phi_g(p),t)$ on $\overline{P}$. We get the
principal $G $-bundle $\overline{\pi}:\overline{P}:=P\times
\mathbb{R}\rightarrow X$. From $\mathcal{A}_t\in\mathcal{C}onn(P)$ and
$\mathcal{A}_0\in\mathcal{F}_G(P,\mathfrak{g})$, we can construct the $1$-form
$\mathcal{C}\in\Omega^1(\overline{P},\mathfrak{g})$
\[
\mathcal{C}:=\tau^*\mathcal{A}_t+\tau^*\mathcal{A}_0\wedge
(\tau_2\circ\overline{\pi})^*dt,
\]
where $\tau :\overline{P}\rightarrow P$ is the natural projection and $\tau_2:X
\rightarrow \mathbb{R}$ is the projection on the second factor. One can check
that $\mathcal{C}$ is a connection on $\overline{P}$ since
$\tau^*\mathcal{A}\in\mathcal{C}onn(\overline{P})$ and
$\tau^*\mathcal{A}_0\wedge
(\tau_2\circ\overline{\pi})^*dt\in\overline{\Omega^1}(\overline{P},\mathfrak{g})$.

Finally we obtain
\[
\textbf{d}^\mathcal{C}\mathcal{C}=\tau^*\mathcal{B}+\tau^*\mathcal{E}\wedge
(\tau_2\circ\overline{\pi})^*dt=:\mathcal{F},
\]
and equations \eqref{YM} are equivalent to the Yang-Mills equations together
with the Bianchi identity (see e.g. \cite{Arms1979, Arms1981},
\cite{ArMaMo1982})
\[
\delta^\mathcal{C}\mathcal{F}=0\;\;\;\text{and}\;\;\;\textbf{d}^\mathcal{C}\mathcal{F}=0.
\]

\medskip \noindent{\bf Hamiltonian formulation.} As in the electromagnetic case, the
configuration space variable is the magnetic potential
$\mathcal{A}\in\mathcal{C}onn(P)$ and the Hamiltonian is
defined on the cotangent bundle $T^*\mathcal{C}onn(P)$ by
\[
H(\mathcal{A},\mathcal{Y})=\frac{1}{2}\int_M\|E\|^2\mu+\frac{1}{2}\int_M\|B\|^2\mu,
\]
where:
\begin{enumerate}
\item[(1)] $E:=\widetilde{\mathcal{E}}\in\Omega^1(M,\operatorname{Ad}P)$ is the
$\operatorname{Ad}P$-valued $1$-form associated, through the map
$\eqref{tilde}$, to the
``electric part"
$\mathcal{E}\in\overline{\Omega^1}(P,\mathfrak{g})$ of the
Yang-Mills field, given by
\[
\mathcal{E}:=-\mathcal{Y}\in\overline{\Omega^1}(P,\mathfrak{g}),
\]
\item[(2)] $B:=\widetilde{\mathcal{B}}\in\Omega^2(M,\operatorname{Ad}P)$ is the
$\operatorname{Ad}P$-valued $2$-form associated, through the map
$\eqref{tilde}$, to the ``magnetic part"
$\mathcal{B}\in\overline{\Omega^2}(P,\mathfrak{g})$ of the
Yang-Mills field, given by the curvature
\[
\mathcal{B}:=\textbf{d}^\mathcal{A}\mathcal{A}\in\overline{\Omega^2}(P,\mathfrak{g}).
\]
\end{enumerate}

As before, we identify the cotangent bundle of $\mathcal{C}onn(P)$ with the
tangent bundle, using the $L^2$ pairing \eqref{L^2pairing}.

Hamilton's equations associated to $H$ are
\[
\displaystyle\frac{\partial \mathcal{B}}{\partial
t}=-\textbf{d}^\mathcal{A}\mathcal{E}\quad\text{and}\quad\frac{\partial
\mathcal{E}}{\partial t}=\delta^\mathcal{A} \mathcal{B},
\]
and the Bianchi identity gives
\[
\textbf{d}^\mathcal{A}\mathcal{B}=0.
\]
To obtain the last equation we use the invariance of the Hamiltonian under
gauge transformations. The action of $\varphi\in\mathcal{G}au(P)$ on
$\mathcal{A}\in\mathcal{C}onn(P)$ is  $\varphi^*\mathcal{A}$ and the
cotangent lift of this action is 
$(\varphi^*\mathcal{A},\varphi^*\mathcal{Y})$. Under this action, $\mathcal{E}$
and $\mathcal{B}$ are transformed into $\varphi^*\mathcal{E}$ and
$\varphi^*\mathcal{B}$, so $H$ is gauge-invariant. The momentum mapping
associated to this Hamiltonian action is 
\[
\textbf{J} :
T^*\mathcal{C}onn(P)\rightarrow\mathfrak{gau}(P)^* \simeq
\mathfrak{gau}(P),\qquad
\textbf{J}(\mathcal{A},\mathcal{Y})=\sigma(\delta^\mathcal{A}\mathcal{Y}),
\]
so the conservation law $\textbf{J}(\mathcal{A},\mathcal{Y})=0$ gives the last
equation
\[
\delta^\mathcal{A}\mathcal{E}=0.
\]
Note that we identify $\mathfrak{gau}(P)^*$ with $\mathfrak{gau}(P)$ via the
$L^2$ pairing \eqref{L^2pairing}.

\section{Equations for the particles}\label{Particles}

We consider the evolution of a non relativistic Yang-Mills charged
particle of mass $m$ in a given Yang-Mills field.

Fix a connection $\mathcal{A}\in\mathcal{C}onn(P)$ and an
equivariant function $\mathcal{A}_0\in \mathcal{F}_G(P,\mathfrak{g})$. The
Yang-Mills field is given by its electric part
$\mathcal{E}=\textbf{d}^\mathcal{A}\mathcal{A}_0$ and its magnetic part
$\mathcal{B}=\textbf{d}^\mathcal{A}\mathcal{A}$.

Consider the right-invariant Lagrangian $L : TP\rightarrow
\mathbb{R}$, given by
\[
L(u_p)=\frac{m}{2}g_{[p]}(T\pi(u_p),T\pi(u_p))+\frac{1}{2}\gamma(\mathcal{A}(u_p)+\mathcal{A}_0(p),\mathcal{A}(u_p)+\mathcal{A}_0(p)).
\]
Since $L$ is $G$-invariant, it induces a Lagrangian on $(TP)/G$. We use the
identification of $(TP)/G$ with $TM\oplus \operatorname{Ad}P$ through the
connection dependent vector bundle isomorphism (see \cite{CeMaRa2001})
\[
\Psi_\mathcal{A} : (TP)/G\rightarrow TM\oplus
\operatorname{Ad}P,\qquad
\Psi_\mathcal{A}([u_p]):=(T_p\pi(u_p),[p,\mathcal{A}(u_p)]_G).
\]
The reduced Lagrangian on $TM\oplus \operatorname{Ad}P$ is given by
\[
l(v_x,\xi_x)=\frac{m}{2}g_x(v_x,v_x)+\frac{1}{2}\gamma_x(\xi_x+A_0(x),\xi_x+A_0(x)),
\]
where $A_0\in\Gamma(\operatorname{Ad}P)$ is associated to $\mathcal{A}_0$ via
the map \eqref{tilde}. By Lagrangian reduction (see \cite{CeMaRa2001}),
$p(t)\in
P$ is a solution of the
Euler-Lagrange equations for $L$ if and only if $x(t):=\pi(p(t))\in M$ and
$\xi(t):=[p(t),\mathcal{A}(\dot{p}(t))]_G\in\operatorname{Ad}P_{x(t)}$ are
solutions of the Lagrange-Poincar\'e equations
\[
\left\lbrace
\begin{array}{ll}
\displaystyle\frac{\partial l}{\partial
x}(\dot{x},\xi)-\frac{D^g}{dt}\frac{\partial l}{\partial
v}(\dot{x},\xi)=\left\langle\frac{\partial l}{\partial
\xi}(\dot{x},\xi),B(\dot{x},.\,)\right\rangle\vspace{0.2cm}\\
\displaystyle\frac{D^\mathcal{A}}{dt}\frac{\partial l}{\partial
\xi}(\dot{x},\xi)=-\operatorname{ad}^*_{\xi}\frac{\partial l}{\partial
\xi}(\dot{x},\xi),
\end{array} \right.
\]
where $D^g/dt$ and $D^\mathcal{A}/dt$ denote the covariant derivatives induced
by $g$ on $T^*M$ and by $\mathcal{A}$ on $(\operatorname{Ad}P)^*$,
respectively,
\[
\frac{\partial l}{\partial v}(v_x,\xi_x)\in T_x^*M \qquad \text{and}\qquad
\frac{\partial l}{\partial \xi}(v_x,\xi_x)\in(\operatorname{Ad}P)^*_x
\]
are usual fiber derivatives of $l $ in the vector bundles $TM $ and
$\operatorname{Ad} P $, and
\[
\frac{\partial l}{\partial x}(v_x,\xi_x) \in T_x^*M
\]
is the partial covariant derivative of $l $ relative to the Levi-Civita
connection on $M $ and the principal connection $\mathcal{A}$ on $P $.
See \cite{CeMaRa2001} for details regarding the Lagrange-Poincar\'e equations.

In terms of the functional derivatives
\[
\frac{\delta l}{\delta x}(v_x,\xi_x), \quad \frac{\delta l}{\delta
v}(v_x,\xi_x)\in
T_xM,\qquad \text{and}\qquad \frac{\delta l}{\delta
\xi}(v_x,\xi_x)\in(\operatorname{Ad}P)_x,
\]
defined similarly, the Lagrange-Poincar\'e equations become
\[
\left\lbrace
\begin{array}{ll}
\displaystyle\frac{\delta l}{\delta x}(\dot{x},\xi)-\frac{D^g}{dt}\frac{\delta
l}{\delta v}(\dot{x},\xi)=\gamma_x\left(\frac{\delta l}{\delta
\xi}(\dot{x},\xi),B(\dot{x},\cdot )\right)^\sharp\vspace{0.2cm}\\
\displaystyle\frac{D^\mathcal{A}}{dt}\frac{\delta l}{\delta
\xi}(\dot{x},\xi)=\left[\xi,\frac{\delta l}{\delta \xi}(\dot{x},\xi)\right]_x,
\end{array} \right.
\]
where $[ \cdot , \cdot ]_x$ is the bracket of elements in
$\left(\operatorname{Ad} P\right)_x$ and  $\frac{D^g}{dt}$ and
$\frac{D^\mathcal{A}}{dt}$ denote the covariant
derivatives on $TM$ and $\operatorname{Ad}P$, respectively. Using that
\begin{align*}
&\frac{\delta l}{\delta x}(v_x,\xi_x)=\frac{\delta \overline{l}}{\delta
x}(v_x,\xi_x)+\gamma_x(\xi_x+A_0(x),\textbf{d}^\mathcal{A}A_0(.\,))^\sharp\\
&\frac{\delta l}{\delta v}(v_x,\xi_x)=\frac{\delta \overline{l}}{\delta
v}(v_x,\xi_x)\\
&\frac{\delta l}{\delta \xi}(v_x,\xi_x)=\xi_x+A_0(x)\\
&\frac{\delta \overline{l}}{\delta
x}(\dot{x},\xi)-\frac{D^g}{dt}\frac{\delta \overline{l}}{\delta
v}(\dot{x},\xi)=-\frac{D^g}{dt}\dot{x}(t),
\end{align*}
where
\[
\overline{l}(v_x,\xi_x)=\frac{m}{2}g_x(v_x,v_x),
\]
we obtain that the trajectory $x(t):=\pi(p(t))\in
M$ and the charge
\[
q(t)=[p(t),\mathcal{A}(\dot{p}(t))+\mathcal{A}_0(p(t))]_G=\frac{\delta
l}{\delta
\xi}(\dot{x}(t),\xi(t))\in(\operatorname{Ad}P)_{x(t)},
\]
are solutions of
\[
\left\lbrace
\begin{array}{ll}
\displaystyle m\frac{D^g}{dt}\dot{x}(t)=\gamma_{x(t)}(q(t),E(\cdot )+B(\cdot
,\dot{x}(t)))^\sharp\vspace{0.2cm}\\
\displaystyle\frac{D^\mathcal{A}}{dt}q(t)+[A_0(x(t)),q(t)]_{x(t)}=0.
\end{array} \right.
\]
The first line is the nonabelian Lorentz equation and the second line
represents
the covariant conservation of charge. These equations are the non-relativistic
Wong equations (see \cite{Wong1970}, \cite{Mo1984}, \cite{CeMaRa2001}).
\medskip

In the case of the trivial $S^1$-bundle $P=M\times S^1$, recall that
$\operatorname{Ad}P=M\times\mathbb{R}$. The Lagrangian is
\[
L(v_x,\theta,\dot{\theta})=\frac{m}{2}g_x(v_x,v_x)+\frac{1}{2}(A(x)(v_x)+\dot{\theta}+A_0(x))^2,
\]
where $A:=\overline{\mathcal{A}}\in\Omega^1(M)$ and
$A_0:=\widetilde{\mathcal{A}_0}$. We have
$(TP)/S^1=TM\oplus\operatorname{Ad}P=TM\times\mathbb{R}$, and
$\Psi_\mathcal{A}(v_x,\dot{\theta})=(v_x,A(x)(v_x)+\dot{\theta})$. So the
reduced Lagrangian is 
\[
l(v_x,\xi)=\frac{m}{2}g_x(v_x,v_x)+\frac{1}{2}(\xi+A_0(x))^2.
\]
By Lagrange-Poincar\'e reduction (\cite{CeMaRa2001}), we obtain that
$p(t)=(x(t),\theta(t))$ is a
solution of the Euler-Lagrange equations if and only if
\[
\left\lbrace
\begin{array}{ll}
\displaystyle m\frac{D^g}{dt}\dot{x}(t)=q(t)(E(\cdot )+B(\cdot
,\dot{x}(t)))^\sharp\vspace{0.2cm}\\
\displaystyle\frac{d}{dt}q(t)=0,
\end{array} \right.
\]
where $q(t):=A(\dot{x}(t))+\dot{\theta}(t)+A_0(x(t))$ is the charge. If $\dim M
= 3 $, in terms of the vector fields $\textbf{E}:=E^\sharp$ and
$\textbf{B}=(\star B)^\sharp$ and using that the charge $q(t)=q$ is conserved,
the previous system becomes simply the Lorentz force law
\[
m\frac{D^g}{dt}\dot{x}(t)=q(\textbf{E}+\dot{x}(t)\times\textbf{B}),
\]
describing the motion of a charged particle of mass $m$ in a fixed
electromagnetic field.

We remark that when the variable $\mathcal{A}_0$ is absent, the Lagrangian is
given
by the Kaluza-Klein metric,
\[
L(u_p)=\frac{1}{2}K_\mathcal{A}(p)(u_p,u_p)=\frac{m}{2}g_{[p]}(T_p\pi(u_p),T\pi_p(u_p))+\gamma(\mathcal{A}(u_p),\mathcal{A}(u_p)).
\]
In this case the Legendre transformation is invertible and the corresponding
Hamiltonian on $T^*P$ is 
\[
H(\alpha_p)=\frac{1}{2}K^*_\mathcal{A}(p)(\alpha_p,\alpha_p),
\]
where $K_\mathcal{A}^*$ is the dual metric on $T^*P$, defined by
\[
K^*_\mathcal{A}(p)\left(K_\mathcal{A}(p)(u_p,\cdot),K_\mathcal{A}(p)(v_p,\cdot)\right):=K_\mathcal{A}(p)(u_p,v_p).
\]

\section{Lagrangian formulation of Euler-Yang-Mills}\label{Lagrangian}

We begin by recalling some facts about Euler-Poincar\'e reduction for
semidirect products (see \cite{HoMaRa1998}, \cite{CeHoHoMa1998},
\cite{CeMaRa2001}). Let $\rho
: G\rightarrow \operatorname{Aut}(V)$ denote a \textit{right\/} Lie group
representation of $G$ in the vector space $V$. As a set, the semidirect product
$S=G\,\circledS\, V$ is the Cartesian product $S=G\times V$ whose group
multiplication is given by
\[
(g_1,v_1)(g_2,v_2)=(g_1g_2,v_2+\rho_{g_2}(v_1)).
\]
The Lie algebra of $S$ is the semidirect product Lie-algebra,
$\mathfrak{s}=\mathfrak{g}\,\circledS\,V$, whose bracket has the expression
\[
\operatorname{ad}_{(\xi_1,v_1)}(\xi_2,v_2)=[(\xi_1,v_1),(\xi_2,v_2)]=([\xi_1,\xi_2],v_1\xi_2-v_2\xi_1),
\]
where $v\xi$ denotes the induced action of $\mathfrak{g}$ on $V$, that is,
\[
v\xi:=\left.\frac{d}{dt}\right|_{t=0}\rho_{\operatorname{exp}(t\xi)}(v)\in
V.
\]
From the expression for the Lie bracket, it follows that for
$(\xi,v)\in\mathfrak{s}$ and $(\mu,a)\in\mathfrak{s}^*$ we have
\[
\operatorname{ad}^*_{(\xi,v)}(\mu,a)=(\operatorname{ad}^*_\xi\mu+v\diamond
a,a\xi),
\]
where $a\xi\in V^*$ and $v\diamond a\in\mathfrak{g}^*$ are given, respectively,
by
\[
a\xi:=\left.\frac{d}{dt}\right|_{t=0}\rho^*_{\operatorname{exp}(-t\xi)}(a)\quad\text{and}\quad
\langle v\diamond a,\xi\rangle_\mathfrak{g}:=-\langle a\xi,v\rangle_V,
\]
where $\left\langle\cdot , \cdot \right\rangle_ \mathfrak{g}: \mathfrak{g}
^\ast
\times \mathfrak{g}\rightarrow \mathbb{R}$ and $\left\langle \cdot , \cdot
\right\rangle_V: V ^\ast \times V \rightarrow \mathbb{R}$ are the duality
parings.
\medskip

\medskip \noindent{\bf Lagrangian semidirect product theory with parameter.}
\begin{itemize}
\item Let $Q$ be a manifold on which $G$ acts trivially and assume that we have
a function $L:TG\times TQ\times V^*\rightarrow\mathbb{R}$ which is right
$G$-invariant.
\item In particular, if $a_0\in V^*$, define the Lagrangian $L_{a_0}:TG\times
TQ\rightarrow\mathbb{R}$ by $L_{a_0}(v_g,u_q):=L(v_g,u_q,a_0)$. Then $L_{a_0}$
is right invariant under the lift to $TG\times TQ$ of the right action of
$G_{a_0}$ on $G\times Q$, where $G_{a_0}$ is the isotropy group of $a_0$.
\item Right $G$-invariance of $L$ permits us to define $l:\mathfrak{g}\times
TQ\times V^*\rightarrow\mathbb{R}$ by
\[
l(T_gR_{g^{-1}}(v_g),u_q,\rho^*_{g}(a_0))=L(v_g,u_q,a_0).
\]
\item For a curve $g(t)\in G$, let $\xi(t):=TR_{g(t)^{-1}}(\dot{g}(t))$ and
define the curve $a(t)$ as the unique solution of the linear differential
equation with time dependent coefficients $\dot{a}(t)=-a(t)\xi(t)$ with initial
condition $a(0)=a_0$. Its solution can be written as $a(t)=\rho^*_{g(t)}(a_0)$.
\end{itemize}

\begin{theorem}\label{EPSD} The following are equivalent:
\begin{itemize}
\item[\bf{i}] Hamilton's variational principle holds:
\[
\delta\int_{t_1}^{t_2}L_{a_0}(g(t),\dot{g}(t),q(t),\dot{q}(t))dt=0,
\]
for variations of $g$ and $q$ with fixed endpoints.
\item[\bf{ii}] $(g(t),q(t))$ satisfies the Euler-Lagrange equations for
$L_{a_0}$ on $G\times Q$.
\item[\bf{iii}] The constrained variational principle
\[
\delta\int_{t_1}^{t_2}l(\xi(t),q(t),\dot{q}(t),a(t))dt=0,
\]
holds on $\mathfrak{g}\times Q$, upon using variations of the form
\[
\delta\xi=\frac{\partial\eta}{\partial t}-[\xi,\eta],\quad \delta a=-a\eta,
\]
where $\eta(t)\in\mathfrak{g}$ vanishes at the endpoints and $\delta q(t)$ is
unrestricted except for vanishing at the endpoints.
\item[\bf{iv}] The following system of Euler-Poincar\'e equations (with a
parameter) coupled with Euler-Lagrange equations holds on $\mathfrak{g}\times
TQ\times V^*$:
\begin{equation}\label{EP}
\frac{\partial}{\partial t}\frac{\delta
l}{\delta\xi}=-\operatorname{ad}^*_\xi\frac{\delta l}{\delta\xi}+\frac{\delta
l}{\delta a}\diamond a,
\end{equation}
and
\[
\frac{\partial}{\partial t}\frac{\partial l}{\partial \dot{q}}-\frac{\partial
l}{\partial q}=0.
\]
\end{itemize}
\end{theorem}
Note that the Euler-Poincar\'e equation \eqref{EP} can be written, in weak
form,
as
\begin{equation}\label{EPweak}
\frac{d}{dt}{\bf D}l(\xi)(\eta)=-{\bf D}l(\xi)([\xi,\eta])+\left\langle
\frac{\delta l}{\delta a}\diamond
a,\eta\right\rangle_\mathfrak{g},\quad \text{for all} \quad \eta\in\mathfrak{g},
\end{equation}
where ${\bf D}$ denotes the Fr\'echet derivative. This formulation will be
useful below.

\medskip \noindent{\bf Ideal compressible adiabatic fluid.} Before treating the Yang-Mills
fluid, we apply the preceding theory to the case of the compressible adiabatic
fluid. For this particular case we choose $G=\mathcal{D}(M)$ and
$V=\mathcal{F}(M)\,\times\,\mathcal{F}(M)$ (in this case $Q$ is absent). We
identify the dual
$\mathcal{F}(M)^*$ with $\mathcal{F}(M)$ via the natural $L^2$ pairing. The
action of $\eta\in\mathcal{D}(M)$ on $(\rho,s)\in V^*$ is 
\[
(\rho,s)\mapsto ((J\eta)(\rho\circ\eta),s\circ\eta),
\]
where $J\eta$ is the Jacobian determinant of $\eta$, $\rho$ is the density of
the fluid and  $s$ is its specific entropy. As usual, we treat the mass density
$\rho$ as a density on $M $ and the entropy $s$ as a function on
$M $; this is why in the previous formula the action of the diffeomorphism
group
is different on the two components.

The Lagrangian is given by
\begin{equation}\label{ICIFLagrangian}
L_{(\rho,s)}(u_\eta)=\frac{1}{2}\int_M\rho g(u_\eta,u_\eta)\mu-\int_M\rho
e(\rho (J\eta)^{-1},s)\mu,
\end{equation}
where $e$ is the fluid's specific internal energy. Application of part
\textbf{iv} in Theorem \ref{EPSD} gives the equations of motion
\begin{equation}\label{ICIF}
\left\lbrace
\begin{array}{ll}
\vspace{0.2cm}\displaystyle\frac{\partial v}{\partial
t}+\nabla_vv=-\frac{1}{\rho}\operatorname{grad}p,\\
\vspace{0.2cm}\displaystyle\frac{\partial \rho}{\partial
t}+\operatorname{div}(\rho v)=0,\\
\displaystyle\frac{\partial s}{\partial t}+\textbf{d}s(v)=0,
\end{array} \right.
\end{equation}
where the pressure is given by $\displaystyle p=\rho^2\frac{\partial
e}{\partial\rho}(\rho,s)$.

\subsection{Yang-Mills ideal fluid}

In the case of the Yang-Mills fluid we choose $G=\mathcal{A}ut(P)$,
$Q=\mathcal{F}_G(P,\mathfrak{g})\times\mathcal{C}onn(P)$ and
$V=\mathcal{F}(M) \times \mathcal{F}(M)$. As before, we use the notations
$\varphi\in\mathcal{A}ut(P)$,
$(\mathcal{A}_0,\mathcal{A})\in\mathcal{F}_G(P,\mathfrak{g})\times\mathcal{C}onn(P)$,
and $(\rho,s)\in V^*$. The action of $\varphi$ on $(\rho,s)$ is given
by
\[
(\rho,s)\mapsto
((J\overline{\varphi})(\rho\circ\overline{\varphi}),s\circ\overline{\varphi}),
\]
where $\overline{ \varphi} \in \mathcal{D}(M)$ is the map induced on the base
$M$ by $\varphi$.
From the expressions of the Lagrangian \eqref{ICIFLagrangian} and of the
Lagrangians for the fields and particles given in \S\ref{Fields} and
\S\ref{Particles}, it follows that the Lagrangian for the Yang-Mills ideal
fluid is defined on the tangent bundle
$T(\mathcal{A}ut(P)\times\mathcal{F}_G(P,\mathfrak{g})\times\mathcal{C}onn(P))$ by
\begin{align}\label{EYMLagrangian}
L_{(\rho,s)}(U_\psi,\mathcal{A}_0,\dot{\mathcal{A}}_0,\mathcal{A},\dot{\mathcal{A}})
&=\frac{1}{2}\int_M\rho g([U_\psi],[U_\psi])\mu  \\
& \qquad +\frac{1}{2}\int_M\rho\|\left(
\mathcal{A}(U_\psi)+\mathcal{A}_0\circ\psi \right) \widetilde{\,}\,\|^2\mu
\nonumber\\
&\quad\quad-\int_M\rho e(\rho (J\overline{\psi})^{-1},s)\mu\nonumber\\
&\quad\quad+\frac{1}{2}\int_M\|E\|^2\mu-\frac{1}{2}\int_M\|B\|^2\mu,
\nonumber
\end{align}
where $[U_\psi]\in T_{\overline{\psi}}\mathcal{D}(M)$ is such that $T\pi\circ
U_\psi=[U_\psi]\circ\pi$. Note that
$\mathcal{A}(U_\psi)+\mathcal{A}_0\circ\psi\in\mathcal{F}_G(P,\mathfrak{g})$,
so
we can consider the section $\left(\mathcal{A}(U_\psi)+\mathcal{A}_0\circ\psi
\right)\widetilde{\,} \in\Gamma(\operatorname{Ad}P)$ and its $L^2$ norm
$\|\left( \mathcal{A}(U_\psi)+\mathcal{A}_0\circ\psi \right) \widetilde{\,}
\,\|$ relative to the Riemannian metric $\gamma_x$. The two last terms of
\eqref{EYMLagrangian} are given as in the Lagrangian \eqref{YMLagrangian}.
Roughly speaking, this Lagrangian has the following structure
\begin{align*} 
& \left\lbrace
\begin{array}{cc}
\text{Integration of the Lagrangian}\\
\text{for the particles}
\end{array} \right\rbrace-\{\text{Internal
energy}\} \\
 & \qquad \qquad +\left\lbrace\begin{array}{cc}
\text{Lagrangian for the}\\
\text{Yang-Mills fields}
\end{array} \right\rbrace.
\end{align*}

Note that $L$ verifies the invariance property needed for an application of
Theorem \ref{EPSD}, that is, $L$ is invariant under the right action of
$\varphi\in\mathcal{A}ut(P)$
\[
(U_\psi,\rho,s)\mapsto
(U_\psi\circ\varphi,(J\overline{\varphi})(\rho\circ\overline{\varphi}),s\circ\overline{\varphi}).
\]
Indeed, we have $[U_\psi\circ\varphi]=[U_\psi]\circ\overline{\varphi}$, so the
invariance of the first term follows by a change of variable in the integral.
The invariance of the second integral follows from the fact that
$\mathcal{A}(U_\psi)+\mathcal{A}_0\circ\psi\in\mathcal{F}_G(P,\mathfrak{g})$
and
that for $f,g\in\mathcal{F}_G(P,\mathfrak{g})$ and $\varphi\in\mathcal{A}ut(P)$
we have
\[
\gamma(\widetilde{f\circ\varphi},\widetilde{g\circ\varphi})=\gamma(\widetilde{f},\widetilde{g})\circ\overline{\varphi},
\]
as functions on $M$.

The reduced Lagrangian $l$ on $\mathfrak{aut}(P)\times
T(\mathcal{F}_G(P,\mathfrak{g})\times\mathcal{C}onn(P))\times
(\mathcal{F}(M)^{*} \times \mathcal{F}(M)^*)$ has the expression 
\begin{align}\label{reducedEYMLagrangian}
& l(U,\mathcal{A}_0,\dot{\mathcal{A}}_0,\mathcal{A},\dot{\mathcal{A}},\rho,s)=\frac{1}{2}\int_M\rho
g([U],[U])\mu \\ 
& \qquad  \qquad +\frac{1}{2}\int_M\rho\|\left( \mathcal{A}(U) 
+\mathcal{A}_0 \right) \widetilde{\,}\,\|^2\mu
-\int_M\rho e(\rho,s)\mu  \nonumber \\
& \qquad \qquad +\frac{1}{2}\int_M\|E\|^2\mu-\frac{1}{2}\int_M\|B\|^2\mu
\nonumber
\end{align}
and the Euler-Poincar\'e equations in weak form are
\begin{equation}\label{Euler-Poincare}
\frac{\partial }{\partial t}{\bf D}l(U)(V)=-{\bf D}l(U)([U,V]_L)+\left\langle
\frac{\delta l}{\delta
(\rho,s)}\diamond(\rho,s),V\right\rangle,
\end{equation}
for all $V\in\mathfrak{aut}(P)$. We now compute these equations.
\medskip 

Recall that the (left) Lie bracket on the Lie algebra $\mathfrak{aut}(P)$ is
\[
[U,V]_L=\operatorname{ad}_UV=-[U,V]_{JL},
\]
where $[\;,\,]_{JL}$ denotes the usual Jacobi-Lie bracket of vector fields. The
following lemma gives the decomposition of $[U,V]_L$ into the horizontal and
vertical parts.

\begin{lemma} Let $\mathcal{A}$ be a connection on the principal bundle $P$ and
let $U,V\in\mathfrak{aut}(P)$. Then we have
\begin{align}\label{bigformula}
[U,V]_L&=\sigma\Big([\mathcal{A}(U),\mathcal{A}(V)]
+\mathbf{d}^\mathcal{A}(\mathcal{A}(U))(V)  \\
& \qquad  -\mathbf{d}^\mathcal{A}(\mathcal{A}(V))(U)+\mathcal{B}(U,V)\Big) +\operatorname{hor}([U,V]_L), \nonumber
\end{align}
where $\operatorname{hor}$ denotes the horizontal part relative to the
connection $\mathcal{A}$. In particular we have the equality
\[
\mathcal{A}\left([U,V]_L\right)=[\mathcal{A}(U),\mathcal{A}(V)]+\mathbf{d}^\mathcal{A}(\mathcal{A}(U))(V)-\mathbf{d}^\mathcal{A}(\mathcal{A}(V))(U)+\mathcal{B}(U,V).
\]
\end{lemma}
\textbf{Proof.} First note that using the Cartan Structure Equations and the
fact that $\mathcal{B}\in\overline{\Omega^2}(P,\mathfrak{g})$, we have
\[
\mathbf{d}\mathcal{A}(\operatorname{hor}U,\sigma(\mathcal{A}(V)))=\mathcal{B}(\operatorname{hor}U,\sigma(\mathcal{A}(V)))-[\mathcal{A}(\operatorname{hor}U),\mathcal{A}(\sigma(\mathcal{A}(V)))]=0.
\]
We also have
\begin{align*}
\mathbf{d}\mathcal{A}(\operatorname{hor}U,\sigma(\mathcal{A}(V)))&=\mathbf{d}(\mathcal{A}(\sigma(\mathcal{A}(V))))(\operatorname{hor}U)-\mathbf{d}(\mathcal{A}(\operatorname{hor}U))(\sigma(\mathcal{A}(V)))\\
&\qquad\qquad-\mathcal{A}([\operatorname{hor}U,\sigma(\mathcal{A}(V))]_{JL})\\
&=\mathbf{d}(\mathcal{A}(V))(\operatorname{hor}U)+\mathcal{A}([\operatorname{hor}U,\sigma(\mathcal{A}(V))]_L).
\end{align*}
These formulas prove that
\begin{equation}\label{dAV(horU)}
\mathbf{d}(\mathcal{A}(V))(\operatorname{hor}U)=-\mathcal{A}([\operatorname{hor}U,\sigma(\mathcal{A}(V))]_L).
\end{equation}
We now compute the Lie bracket $[U,V]_L$. By decomposing $U$ and $V$ into their
vertical and horizontal parts, that is, we write
$U=\sigma(\mathcal{A}(U))+\operatorname{hor}U$ and
$V=\sigma(\mathcal{A}(V))+\operatorname{hor}V$, we obtain four terms. The first
term is
\begin{align*}
[\operatorname{hor}U,\operatorname{hor}V]_L&=\sigma\left(\mathcal{A}\left([\operatorname{hor}U,\operatorname{hor}V]_L\right)\right)+\operatorname{hor}[\operatorname{hor}U,\operatorname{hor}V]_L\\
&=\sigma(\mathcal{B}(U,V))+\operatorname{hor}[U,V]_L,
\end{align*}
where we used the equalities
\[
\mathcal{B}(U,V)=-\mathcal{A}([\operatorname{hor}U,\operatorname{hor}V]_{JL})
\]
and
\[
T\pi\circ [U,V]_L=[[U],[V]]_L \circ \pi.
\]
Since $[\sigma(\mathcal{A}(U)),\operatorname{hor}V]_L $ is vertical (apply the
formula above), the second term is
\begin{align*} 
[\sigma(\mathcal{A}(U)),\operatorname{hor}V]_L
& =\sigma(\mathcal{A}([\sigma(\mathcal{A}(U)),\operatorname{hor}V]_L))=\mathbf{d}(\mathcal{A}(U))(\operatorname{hor}V) \\
& =\mathbf{d}^\mathcal{A}(\mathcal{A}(U))(V),
\end{align*} 
by formula \eqref{dAV(horU)}. There is an analogous formula for the third term
$[\operatorname{hor}U,\sigma(\mathcal{A}(V))]_L$. Using the Lie algebra
isomorphism $\sigma : \mathcal{F}_G(P,\mathfrak{g})\rightarrow\mathfrak{gau}(P)$
defined in \eqref{sigma}, the fourth term is
\[
[\sigma(\mathcal{A}(U)),\sigma(\mathcal{A}(V))]=\sigma([\mathcal{A}(U),\mathcal{A}(V)]_L).
\]
Summing these four terms we obtain the desired formula
\eqref{bigformula}.$\qquad\blacksquare$

\medskip

Inspired by the Kaluza-Klein metric \eqref{KK_metric}, we define on
$\mathfrak{aut}(P)$ a non-degenerate bilinear form given by
\[
\langle
U,V\rangle_\mathcal{A}:=\int_Mg([U],[V]) \mu +\int_M\gamma\left(\widetilde{\mathcal{A}(U)},\widetilde{\mathcal{A}(V)}\right)\mu.
\]
Therefore we have
\begin{align}\label{ad^*}
&\langle W,[U,V]_L\rangle_\mathcal{A} 
=\int_Mg([W],[[U,V]_L])\mu+\int_M\gamma\left(\widetilde{\mathcal{A}(W)},\widetilde{\mathcal{A}([U,V]_L)}\right)\mu \\
&\quad  =\int_Mg([W],[[U],[V]]_L)\mu+\int_M\gamma\left(\widetilde{\mathcal{A}(W)},\widetilde{[\mathcal{A}(U),\mathcal{A}(V)]}\right)\mu\nonumber\\
&\quad \quad +\int_M\gamma\left(\widetilde{\mathcal{A}(W)},\widetilde{\textbf{d}^\mathcal{A}(\mathcal{A}(U))}[V]\right)\mu \nonumber \\
& \quad \quad -\int_M\gamma\left(\widetilde{\mathcal{A}(W)},
\widetilde{\textbf{d}^\mathcal{A}(\mathcal{A}(V))}[U]\right)\mu\nonumber\\
&\quad \quad  +\int_M\gamma\left(\widetilde{\mathcal{A}(W)},\widetilde{\mathcal{B}}([U],[V])\right)\mu\nonumber\\
& \quad =\int_Mg\left(\operatorname{ad}^\dagger_{[U]}[W]
\phantom{\left(\widetilde{\textbf{d}^\mathcal{A}(\mathcal{A}(U))}\right)^\sharp} \right.  \nonumber \\
&\qquad \qquad \left.  +\; \gamma\left(\widetilde{\mathcal{A}(W)},\widetilde{\textbf{d}^\mathcal{A}(\mathcal{A}(U))}(\cdot
)+\widetilde{\mathcal{B}}([U],\cdot )\right)^\sharp, [V]\right)\mu\nonumber\\
&\quad \quad +\int_M\gamma\left(\widetilde{[\mathcal{A}(W),\mathcal{A}(U)]}+\widetilde{\textbf{d}^\mathcal{A}(\mathcal{A}(W))}[U] \right. 
\nonumber  \\
& \left. \phantom{\widetilde{\textbf{d}^\mathcal{A}(\mathcal{A}(W))}}
+\operatorname{div}([U])\widetilde{\mathcal{A}(W)},\;\widetilde{\mathcal{A}(V)}\right)\mu,
\nonumber
\end{align}
where in the last equality, $\operatorname{ad}^\dagger$ denotes the $L^2$ adjoint
of
$\operatorname{ad}$ relative to the metric $g $, and $\sharp$ is the index
raising operator associated to $g$. Note  that for $u,w\in\mathfrak{X}(M)$,
$\operatorname{ad}^\dagger$ is given by
\begin{equation}\label{ad^dagger}
\operatorname{ad}^\dagger_uw=\nabla_uw+\nabla u^T\cdot w+w\operatorname{div}u.
\end{equation}
In the second summand of the last equality in \eqref{ad^*} we used the
following
Lemma.

\begin{lemma}\label{integrationbypart}
Consider an $\operatorname{Ad}$-invariant inner product $\gamma$ on
$\mathfrak{g}$ and the induced vector bundle metric on $\operatorname{Ad}P$,
also denoted by $\gamma$. Then for $v\in\mathfrak{X}(M)$ and
$f,g\in\mathcal{F}_G(P,\mathfrak{g})$ we have
\[
\int_M\gamma\left(\widetilde{{\bf
d}^\mathcal{A}f}(v),\widetilde{g}\right)\mu=-\int_M\gamma\left(\widetilde{f},\widetilde{{\bf
d}^\mathcal{A}g}(v)\right)\mu-\int_M\gamma(\widetilde{f},\widetilde{g})(\operatorname{div}v)\mu.
\]
\end{lemma}
\textbf{Proof.} One verifies that for any $v \in \mathfrak{X}(M)$ we have
\[
\textbf{d}\left(\gamma(\widetilde{f},\widetilde{g})\right)(v)=\gamma\left(\widetilde{\textbf{d}^\mathcal{A}f}(v),\widetilde{g}\right)+\gamma\left(\widetilde{f},\widetilde{\textbf{d}^\mathcal{A}g}(v)\right).
\]
Integrating over $M$ gives the result. Indeed, denoting by $h$ the real valued
function $\gamma(\widetilde{f},\widetilde{g})$, we obtain
\[
\int_M\textbf{d}h(v) \mu =\int_M\operatorname{div}(hv) \mu -
\int_Mh(\operatorname{div}v) \mu=-\int_Mh(\operatorname{div}v) \mu,
\]
by the divergence Theorem.$\qquad\blacksquare$
\medskip

Using  \eqref{ad^*} and the formula
\begin{align*} 
{\bf D}l(U)(V)&=\frac{1}{2}\int_Mg(\rho
[U],[V])\mu+\frac{1}{2}\int_M\gamma\left(
\rho\left(\mathcal{A}(U)+\mathcal{A}_0
\right)\widetilde{\,}\,,\widetilde{\mathcal{A}(V)}\right)\mu \\
& =\langle \rho
U+\sigma(\mathcal{A}_0),V\rangle_\mathcal{A},
\end{align*} 
we obtain
\begin{align}\label{compressibleformula}
{\bf D}l&(U)([U,V]_L) = \langle \rho
U+\sigma(\mathcal{A}_0),[U,V]_L\rangle_\mathcal{A} \\
&=\int_Mg\left(\operatorname{ad}^\dagger
_{[U]}\rho[U]+\gamma\left(\widetilde{\mathcal{Q}},\widetilde{\textbf{d}^\mathcal{A}(\mathcal{A}(U))}(.)+\widetilde{\mathcal{B}}([U],.)\right)^\sharp,[V]\right)\mu\nonumber\\
&\quad+\int_M\gamma\left(\widetilde{[\mathcal{Q},\mathcal{A}(U)]}+\widetilde{\textbf{d}^\mathcal{A}\mathcal{Q}(U)}+\operatorname{div}([U])\widetilde{\mathcal{Q}},\widetilde{\mathcal{A}(V)}\right) \mu , \nonumber
\end{align}
where
\begin{equation}
\label{def_charge_density}
\mathcal{Q}:=\rho(\mathcal{A}(U)+\mathcal{A}_0)\in\mathcal{F}_G(P,\mathfrak{g})
\end{equation}
is the charge density.

On the other hand we have, using the notations
$A_0:=\widetilde{\mathcal{A}_0}\in\Gamma(\operatorname{Ad}P)$,
$\dot{A}:=\widetilde{\dot{\mathcal{A}}}\in\Omega^1(M,\operatorname{Ad}P)$, and
$Q:=\widetilde{\mathcal{Q}}\in\Gamma(\operatorname{Ad}P)$,
\begin{align*}
&\frac{\partial}{\partial t}{\bf
D}l(U)(V)=\frac{d}{dt}\left[\int_Mg\left(\rho[U],[V]\right)\mu+\int_M\gamma\left(Q,\widetilde{\mathcal{A}(V)}\right)\mu\right]\\
&\quad =\int_Mg\left(\frac{\partial}{\partial
t}\rho[U],[V]\right)\mu+\int_M\gamma\left(\dot{Q},\widetilde{\mathcal{A}(V)}\right)\mu+\int_M\gamma\left(Q,\dot{A}[V]\right)\mu\\
&\quad =\int_Mg\left(\rho\frac{\partial}{\partial
t}[U]-\operatorname{div}(\rho[U])[U]+\gamma\left(Q,\dot{A}(.)\right)^\sharp,[V]\right)\mu \\
& \quad \qquad +\int_M\gamma\left(\dot{Q},\widetilde{\mathcal{A}(V)}\right)\mu,
\end{align*}
where we used the equation $\dot{\rho}=-\operatorname{div}(\rho[U])$. Using the
equalities
\begin{align*}
\frac{\delta l}{\delta
\rho}&=\frac{1}{2}g([U],[U])+\frac{1}{2}\gamma\left(\frac{Q}{\rho},\frac{Q}{\rho}\right)-e-\rho\frac{\partial
e}{\partial\rho},\\
\frac{\delta l}{\delta s}&=-\rho\frac{\partial e}{\partial s},\\
\frac{\delta l}{\delta (\rho,s)}\diamond
(\rho,s)&=\rho\operatorname{grad}\frac{\delta l}{\delta
\rho}-\frac{\delta l}{\delta s}\operatorname{grad}s,\\
&=\rho\nabla[U]^T\cdot
[U]+\frac{1}{2}\rho\operatorname{grad}\gamma\left(\frac{Q}{\rho},\frac{Q}{\rho}\right)-\operatorname{grad}\left(\rho^2\frac{\partial
e}{\partial\rho}\right),
\end{align*}
equation \eqref{Euler-Poincare} yields the system
\begin{equation}\label{system1}
\left\lbrace
\begin{array}{ll}
\displaystyle\rho\frac{\partial}{\partial
t}[U]+\rho\nabla_{[U]}[U]+\gamma\left(Q,\dot{A}(.)+\widetilde{\textbf{d}^\mathcal{A}(\mathcal{A}(U))}(\cdot
)+B([U],\cdot )\right)^\sharp\vspace{0.2cm}\\
\qquad\qquad\qquad\qquad=\displaystyle\frac{1}{2}\rho\operatorname{grad}\gamma\left(\frac{Q}{\rho},\frac{Q}{\rho}\right)-\operatorname{grad}p\vspace{0.2cm}\\
\displaystyle\frac{\partial}{\partial
t}\mathcal{Q}+[\mathcal{Q},\mathcal{A}(U)]+\textbf{d}^\mathcal{A}\mathcal{Q}(U)+\operatorname{div}([U])\mathcal{Q}=0.
\end{array} \right.
\end{equation}
Denoting
$\displaystyle\mathcal{S}:=\frac{\mathcal{Q}}{\rho}=\mathcal{A}(U)+\mathcal{A}_0$,
several applications of Lemma \ref{integrationbypart}, give
\[
\int_M\gamma\left(S,\widetilde{\textbf{d}^\mathcal{A}(\mathcal{A}(U))}(v)\right) \mu =\frac{1}{2}\int_M\textbf{d}(\gamma(S,S))(v) \mu -\int_M\gamma\left(S,\widetilde{\textbf{d}^\mathcal{A}\mathcal{A}_0}(v)\right) \mu ,
\]
for all $v\in\mathfrak{X}(M)$. So we get
\[
\gamma\left(Q,\widetilde{\textbf{d}^\mathcal{A}(\mathcal{A}(U))}(\cdot
)\right)^\sharp=\frac{1}{2}\rho\operatorname{grad}\gamma\left(\frac{Q}{\rho},\frac{Q}{\rho}\right)-\gamma\left(Q,\widetilde{\textbf{d}^\mathcal{A}\mathcal{A}_0}(\cdot
)\right)^\sharp.
\]
With this formula and the equality
\[
[\mathcal{Q},\mathcal{A}(U)]=[\mathcal{A}_0,\mathcal{Q}],
\]
the system \eqref{system1} is equivalent to
\begin{equation}\label{system2}
\left\lbrace
\begin{array}{ll}
\displaystyle\frac{\partial}{\partial
t}[U]+\nabla_{[U]}[U]=\frac{1}{\rho}\gamma\left(Q,-\dot{A}(\cdot
)+\widetilde{\textbf{d}^\mathcal{A}\mathcal{A}_0}(\cdot )+B(\cdot
,[U])\right)^\sharp \vspace{0.2cm} \\
\qquad \qquad \qquad \qquad \qquad \qquad 
\displaystyle -\frac{1}{\rho}\operatorname{grad}p\vspace{0.2cm}\\
\displaystyle\frac{\partial}{\partial
t}\mathcal{Q}+[\mathcal{A}_0,\mathcal{Q}]+\textbf{d}^\mathcal{A}\mathcal{Q}(U)+\operatorname{div}([U])\mathcal{Q}=0,
\end{array} \right.
\end{equation}
which is the same as
\begin{equation}\label{system3}
\left\lbrace
\begin{array}{ll}
\displaystyle\frac{\partial v}{\partial
t}+\nabla_vv=\frac{1}{\rho}\gamma\left(Q,E(\cdot )+B(\cdot
,v)\right)^\sharp-\frac{1}{\rho}\operatorname{grad}p\vspace{0.2cm}\\
\displaystyle\frac{\partial Q}{\partial
t}+[A_0,Q]+\nabla^\mathcal{A}_vQ+Q\operatorname{div}v=0,
\end{array} \right.
\end{equation}
where $v:=[U]\in\mathfrak{X}(M)$ is the Eulerian velocity.

\medskip

We compute now the Euler-Lagrange equations relative to the Lagrangian
\eqref{reducedEYMLagrangian} and the variables $(\mathcal{A}_0,\mathcal{A})$.
The computations are similar to those done in Paragraph 3.3. We find
\[
\frac{\partial
l}{\partial\mathcal{A}_0}=\delta^\mathcal{A}\mathcal{E}+\mathcal{Q},\;\frac{\partial
l}{\partial\dot{\mathcal{A}}_0}=0,\; \frac{\partial
l}{\partial\mathcal{A}}=-\delta^\mathcal{A}\mathcal{B}+[\mathcal{A}_0,\mathcal{E}]+\mathcal{Q}\otimes\pi^*v^\flat,\;
\frac{\partial l}{\partial\dot{\mathcal{A}}}=-\mathcal{E},
\]
where $\mathcal{Q}\,\otimes\,\pi^*v^\flat \in \overline{ \Omega^1}(P,
\mathfrak{g})$ is given by
\[
\left(\mathcal{Q}\otimes\pi^*v^\flat\right)(u_p):=\mathcal{Q}(p)\left(\pi^*v^\flat(u_p)\right)=\mathcal{Q}(p)g_x(v(x),T_p\pi(u_p)),\quad
x=\pi(p).
\]
Note that we have $\widetilde{\mathcal{Q}\,\otimes\,\pi^*v^\flat}=Q\otimes
v^\flat$, the $1$-form on $M$, with
values in $\operatorname{Ad}P$, given by
\[
\left(Q\otimes
v^\flat\right)(u_x)=Q(x)v^\flat(u_x)=Q(x)g_x(v(x),u_x)\in(\operatorname{Ad}P)_x.
\]
For the computation of the partial derivative $\frac{\partial
l}{\partial\mathcal{A}}$, we use the identity
\[
\gamma(Q,\dot{A}(v))=(g\gamma)(Q\otimes v^\flat,\dot{A}).
\]
The resulting Euler-Lagrange equations are 
\[
\delta^\mathcal{A}
\mathcal{E}=-\mathcal{Q}\;\;\;\text{and}\;\;\;\displaystyle\frac{\partial
\mathcal{E}}{\partial t}+[\mathcal{A}_0,\mathcal{E}]=\delta^\mathcal{A}
\mathcal{B}-\mathcal{Q}\otimes\pi^*v^\flat.
\]
As in Paragraph 3.3, the relations
$\mathcal{E}=-\dot{\mathcal{A}}+\textbf{d}^\mathcal{A}\mathcal{A}_0$ and
$\mathcal{B}=\textbf{d}^\mathcal{A}\mathcal{A}$ give the equations
\[
\displaystyle\frac{\partial \mathcal{B}}{\partial
t}+[\mathcal{A}_0,\mathcal{B}]=-\textbf{d}^\mathcal{A}\mathcal{E}\;\;\;\text{and}\;\;\;\textbf{d}^\mathcal{A}\mathcal{B}=0.
\]

Summarizing,  we have proved the following theorem, which is one of the main
results of this paper.

\begin{theorem}\label{THM1} Let $(\psi,\mathcal{A}_0,\mathcal{A})$ be a curve
in
$\mathcal{A}ut(P)\times\mathcal{F}_G(P,\mathfrak{g})\times\mathcal{C}onn(P)$
and
consider the curve
$(U,\mathcal{A}_0,\mathcal{A}):=(\dot{\psi}\circ\psi^{-1},\mathcal{A}_0,\mathcal{A})$
in
$\mathfrak{aut}(P)\times\mathcal{F}_G(P,\mathfrak{g})\times\mathcal{C}onn(P)$.
Then $(\psi,\mathcal{A}_0,\mathcal{A})$ is a solution of the Euler-Lagrange
equations associated to the Lagrangian $L_{(\rho_0,s_0)}$ given in
\eqref{EYMLagrangian} if and only if $(U,\mathcal{A}_0,\mathcal{A})$ is a
solution of the Euler-Yang-Mills equations:
\begin{equation}\label{EYM}
\left\lbrace
\begin{array}{ll}
\displaystyle\frac{\partial v}{\partial
t}+\nabla_vv=\frac{1}{\rho}\gamma\left(Q,E(\cdot )+B(\cdot
,v)\right)^\sharp-\frac{1}{\rho}\operatorname{grad}p,\vspace{0.2cm}\\
\displaystyle\frac{\partial\rho}{\partial t}+\operatorname{div}(\rho
v)=0,\,\rho(0)=\rho_0,\quad\frac{\partial s}{\partial t}+{\bf d}s
(v)=0,\,s(0)=s_0,\vspace{0.2cm}\\
\displaystyle\frac{\partial Q}{\partial
t}+[A_0,Q]+\nabla^\mathcal{A}_vQ+Q\operatorname{div}v=0,\vspace{0.2cm}\\
\displaystyle\displaystyle\frac{\partial \mathcal{E}}{\partial
t}+[\mathcal{A}_0,\mathcal{E}]=\delta^\mathcal{A}
\mathcal{B}-\mathcal{Q}\otimes\pi^*v^\flat,\quad\delta^\mathcal{A}
\mathcal{E}=-\mathcal{Q},\vspace{0.2cm}\\
\displaystyle\frac{\partial \mathcal{B}}{\partial
t}+[\mathcal{A}_0,\mathcal{B}]=-{\bf d}^\mathcal{A}\mathcal{E},\quad{\bf
d}^\mathcal{A}\mathcal{B}=0,
\end{array} \right.
\end{equation}
where
\begin{align*}
p:&=\rho^2\frac{\partial e}{\partial\rho}(\rho,s),\quad
v:=[U]\in\mathfrak{X}(M),\\
\mathcal{E}:&=-\dot{\mathcal{A}}+{\bf
d}^\mathcal{A}\mathcal{A}_0\in\overline{\Omega^1}(P,\mathfrak{g})\quad\text{and}\quad
E:=\widetilde{\mathcal{E}}\in\Omega^1(M,\operatorname{Ad}P),\\
\mathcal{B}:&={\bf d}^\mathcal{A}\mathcal{A}\quad\text{and}\quad
B:=\widetilde{\mathcal{B}},\\
\mathcal{Q}:&=\rho(\mathcal{A}(U)+\mathcal{A}_0)\quad\text{and}\quad
Q=\widetilde{\mathcal{Q}}.
\end{align*}

\end{theorem}

\begin{corollary}\label{corollary_EM} In the case of the trivial bundle
$P=M\times S^1$ and assuming that the fluid is composed of particles of mass
$m$
and charge $q$, we obtain the Euler-Maxwell equations
\begin{equation}\label{EM}
\left\lbrace
\begin{array}{ll}
\vspace{0.2cm}\displaystyle\frac{\partial v}{\partial
t}+\nabla_vv=\frac{q}{m}({\bf E}+v\times {\bf
B})-\frac{1}{\rho}\operatorname{grad}p,\\
\displaystyle\frac{\partial\rho}{\partial t}+\operatorname{div}(\rho
v)=0,\,\rho(0)=\rho_0,\quad\frac{\partial s}{\partial t}+{\bf d}s
(v)=0,\,s(0)=s_0,\vspace{0.2cm}\\
\vspace{0.2cm}\displaystyle\frac{\partial {\bf E}}{\partial
t}=\operatorname{curl}{\bf B}-\frac{q}{m}\rho v,\quad\frac{\partial {\bf
B}}{\partial t}=-\operatorname{curl}{\bf E},\\
\displaystyle\operatorname{div}{\bf
E}=\frac{q}{m}\rho,\quad\operatorname{div}{\bf B}=0,
\end{array} \right.
\end{equation}
where
\[
{\bf E}:=E^\sharp\quad\text{and}\quad{\bf B}:=(\star B)^\sharp.
\]
\end{corollary}
\textbf{Proof. } If we define $Q_t=\rho_t \frac{q_t}{m}$, the equation for $Q$ 
in \eqref{EYM} becomes
\[
0 = \frac{\partial q_t}{\partial t}+\textbf{d}q_t(v)= \frac{d}{dt}q_t(x(t)),
\]
where $x(t) $ is the trajectory of the particle starting at $x(0)$. Since all
particles have the same charge $q \in \mathbb{R}$ by hypothesis, we conclude
that $q(t, x)$ is a constant. Therefore, the equation for $Q $ in \eqref{EYM}
disappears. It is easily seen that  the other equations become the ones in
\eqref{EM}. $\qquad\blacksquare$
\medskip

We end this section by examining more carefully the case of a trivial principal
bundle $P=M\times G$. We use the fact, already pointed out in the introduction,
that in this case the automorphism group is a semidirect product of two groups.

In the trivial bundle case, we have a connection independent $L^2$ pairing on
$\mathfrak{aut}(P)$, given by
\[
\langle(m,\nu),(v,\theta)\rangle=\int_Mg(m,v)\mu+\int_M\gamma(\nu,\theta)\mu.
\]
Using this pairing, the expression \eqref{Liebracket} for the Lie bracket on
the semidirect product Lie algebra, the expression \eqref{ad^dagger}, and
integration by parts, we obtain the following expression for
$\operatorname{ad}^\dagger$:
\begin{align}\label{adjoint^*}
&\operatorname{ad}^\dagger_{(v,\theta)}(m,\nu) \\
& \qquad =\left(\operatorname{ad}^\dagger_vm+\gamma(\nu,\textbf{d}\theta(\cdot))^\sharp,\nu\operatorname{div}v+\textbf{d}\nu(v)
+[\nu,\theta]\right)
\nonumber  \\
& \qquad =\left(\nabla_vm+\nabla v^T\cdot
m+m\operatorname{div}v+\gamma(\nu,\textbf{d}\theta(\cdot
))^\sharp, \right. \nonumber  \\
& \qquad \qquad \qquad 
\left. \phantom{\textbf{d}\theta(\cdot))^\sharp}
\nu\operatorname{div}v+\textbf{d}\nu(v)+[\nu,\theta]\right).
\nonumber
\end{align}
The reduced Lagrangian 
\[
l:\mathfrak{aut}(P)\times
T(\mathcal{F}_G(P,\mathfrak{g})\times\mathcal{C}onn(P))\times(\mathcal{F}(M)^{*}
\times \mathcal{F}(M)^*)\rightarrow\mathbb{R}
\]
is 
\begin{align}
l(v,\theta,\mathcal{A}_0,\dot{\mathcal{A}}_0,\mathcal{A},\dot{\mathcal{A}},\rho,s)&=\frac{1}{2}\int_M\rho
g(v,v)\mu+\frac{1}{2}\int_M\rho\|\overline{\mathcal{A}}(v)+\theta+A_0\|^2\mu\nonumber\\
&\quad\quad-\int_M\rho
e(\rho,s)\mu+\frac{1}{2}\int_M\|E\|^2\mu-\frac{1}{2}\int_M\|B\|^2\mu.
\end{align}
and we have
\[
\frac{\delta l}{\delta
v}=\rho(v+\gamma(\overline{\mathcal{A}}(v)+\theta+A_0,\overline{\mathcal{A}}(\cdot
))^\sharp)\quad\text{and}\quad\frac{\delta l}{\delta
\theta}=\rho\left(\overline{\mathcal{A}}(v)+\theta+A_0\right).
\]
The Euler-Poincar\'e equations are
\[
\frac{\partial}{\partial t}\left(\frac{\delta l}{\delta v},\frac{\delta
l}{\delta
\theta}\right)=-\operatorname{ad}^\dagger_{(v,\theta)}\left(\frac{\delta
l}{\delta v},\frac{\delta l}{\delta \theta}\right)+\frac{\delta l}{\delta
(\rho,s)}\diamond (\rho,s),
\]
and a long direct computation gives, as expected, the system
\begin{equation}
\left\lbrace
\begin{array}{ll}
\displaystyle\frac{\partial v}{\partial
t}+\nabla_vv=\frac{1}{\rho}\gamma\left(Q,E(\cdot )+B(\cdot
,v)\right)^\sharp-\frac{1}{\rho}\operatorname{grad}p,\vspace{0.2cm}\\
\displaystyle\frac{\partial Q}{\partial
t}+[\overline{\mathcal{A}}(v)+A_0,Q]+\textbf{d}Q(v)+Q\operatorname{div}v=0,
\end{array} \right.
\end{equation}
where
\[
Q:=\frac{\delta l}{\delta
\theta}=\rho\left(\overline{\mathcal{A}}(v)+\theta+A_0\right)\in\mathcal{F}(M,\mathfrak{g}),\;
E\in\Omega^1(M,\mathfrak{g}),\; \
B\in\Omega^2(M,\mathfrak{g}).
\]

\subsection{The incompressible and homogeneous case}

In the incompressible case we choose $G=\mathcal{A}ut_\mu(P)$, the Lie group of
all automorphisms $\varphi\in\mathcal{A}ut(P)$ such that
$\overline{\varphi}\in\mathcal{D}_\mu(M)$. Since the fluid is homogeneous, the
advected variables $\rho$ and $s$ are absent. Therefore, we can use
standard
Euler-Poincar\'e reduction with parameters
$(\mathcal{A}_0,\mathcal{A})\in\mathcal{F}_G(P,\mathfrak{g})\times\mathcal{C}onn(P)$
(take $V=0$ in the semidirect theory). The Lagrangian for the incompressible
homogeneous Yang-Mills ideal fluid is defined on the tangent bundle
$T(\mathcal{A}ut_\mu(P)\times\mathcal{F}_G(P,\mathfrak{g})\times\mathcal{C}onn(P))$
and is given by
\begin{align}\label{IEYMLagrangian}
L(U_\psi,\mathcal{A}_0,\dot{\mathcal{A}}_0\mathcal{A},\dot{\mathcal{A}})&=\frac{1}{2}\int_M
g([U_\psi],[U_\psi])\mu \\
& \quad \quad +\frac{1}{2}\int_M\|\left(\mathcal{A}(U_\psi)+\mathcal{A}_0\circ\psi \right)\widetilde{\,}\,\|^2\mu\nonumber\\
&\quad\quad+\frac{1}{2}\int_M\|E\|^2\mu-\frac{1}{2}\int_M\|B\|^2\mu.
\nonumber 
\end{align}
The computations of the Euler-Poincar\'e equations are similar to those done in
the compressible case, except that we have $\operatorname{div}([U])=0$ and we
must replace formula \eqref{compressibleformula} by formula
\begin{align*}
&{\bf D}l(U)([U,V]_L) \\
&\quad =\int_Mg\left(\operatorname{P}_e\left(\operatorname{ad}^\dagger
_{[U]}[U]+\gamma\left(\widetilde{\mathcal{Q}},\widetilde{\textbf{d}^\mathcal{A}(\mathcal{A}(U))}(.)+\widetilde{\mathcal{B}}([U],.)\right)^\sharp\right),[V]\right)\mu\nonumber\\
&\quad\qquad +\int_M\gamma\left(\widetilde{[\mathcal{Q},\mathcal{A}(U)]}+\widetilde{\textbf{d}^\mathcal{A}\mathcal{Q}(U)},\widetilde{\mathcal{A}(V)}\right) \mu ,
\end{align*}
where $\operatorname{P}_e:\mathfrak{X}(M)\rightarrow\mathfrak{X}_{\rm div}(M)$
is
the projector associated to the $L^2$ orthogonal Hodge decomposition
\[
\mathfrak{X}(M)=\mathfrak{X}_{\rm
div}(M)\oplus\operatorname{grad}(\mathcal{F}(M)).
\]
We finally get the following result.

\begin{theorem} Let $(\psi,\mathcal{A}_0,\mathcal{A})$ be a curve in
$\mathcal{A}ut_\mu(P)\times\mathcal{F}_G(P,\mathfrak{g})\times\mathcal{C}onn(P)$
and consider the curve
$(U,\mathcal{A}_0,\mathcal{A}):=(\dot{\psi}\circ\psi^{-1},\mathcal{A}_0,\mathcal{A})$
in
$\mathfrak{aut}_\mu(P)\times\mathcal{F}_G(P,\mathfrak{g})\times\mathcal{C}onn(P)$.
Then $(\psi,\mathcal{A}_0,\mathcal{A})$ is a solution of the Euler-Lagrange
equations associated to the Lagrangian \eqref{IEYMLagrangian} if and only if
$(U,\mathcal{A}_0,\mathcal{A})$ is a solution of the incompressible homogeneous
Euler-Yang-Mills equations:
\begin{equation}\label{EYMI}
\left\lbrace
\begin{array}{ll}
\displaystyle\frac{\partial v}{\partial t}+\nabla_vv=\gamma\left(Q,E(\cdot )+B(
\cdot ,v)\right)^\sharp-\operatorname{grad}p,\vspace{0.2cm}\\
\displaystyle\frac{\partial Q}{\partial
t}+[A_0,Q]+\nabla^\mathcal{A}_vQ=0,\vspace{0.2cm}\\
\displaystyle\displaystyle\frac{\partial \mathcal{E}}{\partial
t}+[\mathcal{A}_0,\mathcal{E}]=\delta^\mathcal{A}
\mathcal{B}-\mathcal{Q}\otimes\pi^*v^\flat,\quad\delta^\mathcal{A}
\mathcal{E}=-\mathcal{Q},\vspace{0.2cm}\\
\displaystyle\frac{\partial \mathcal{B}}{\partial
t}+[\mathcal{A}_0,\mathcal{B}]=-{\bf d}^\mathcal{A}\mathcal{E},\quad{\bf
d}^\mathcal{A}\mathcal{B}=0,
\end{array} \right.
\end{equation}
where
\begin{align*}
v:&=[U]\in\mathfrak{X}_{\rm div}(M),\\
\mathcal{E}:&=-\dot{\mathcal{A}}+{\bf
d}^\mathcal{A}\mathcal{A}_0\in\overline{\Omega^1}(P,\mathfrak{g})\quad\text{and}\quad
E:=\widetilde{\mathcal{E}}\in\Omega^1(M,\operatorname{Ad}P),\\
\mathcal{B}:&={\bf d}^\mathcal{A}\mathcal{A}\quad\text{and}\quad
B:=\widetilde{\mathcal{B}},\\
\mathcal{Q}:&=\mathcal{A}(U)+\mathcal{A}_0\quad\text{and}\quad
Q=\widetilde{\mathcal{Q}}.
\end{align*}
\end{theorem}

Note that the pressure is in this case determined from $v $, $E $, and $B $
through the Green's function of the Laplacian on $M$. This is in contrast to
\eqref{EYM} where the pressure was given by the internal energy.

If $P$ is a trivial bundle, one gets the incompressible homogeneous version of
the Euler-Yang-Mills equations  (corresponding to the group $\mathcal{D}_ \mu
(M) \,\circledS\, \mathcal{F}(M, G)$) by replacing in formula
\eqref{adjoint^*} the vector fields by their projection onto  their divergence
free part, namely, 
\begin{align}\label{compressibleadjoint^*}
\operatorname{ad}^\dagger_{(v,\theta)}(m,\nu)&=\left(\operatorname{ad}^\dagger_vm+\operatorname{P}_e\left(\gamma(\nu,\textbf{d}\theta(\cdot))^\sharp\right),\textbf{d}\nu(v)+[\nu,\theta]\right)\nonumber\\
&=\left(\operatorname{P}_e\left(\nabla_vm+\nabla v^T\cdot
m+\gamma(\nu,\textbf{d}\theta(\cdot))^\sharp\right),\textbf{d}\nu(v)+[\nu,\theta]\right).
\end{align}

One can also adapt our method to the case of the incompressible but
non-homogeneous Yang-Mills fluid. It suffices to apply the semidirect product
theory with $G=\mathcal{A}ut_\mu(P)$,
$Q=\mathcal{F}_G(P,\mathfrak{g})\times\mathcal{C}onn(P)$ and
$V=\mathcal{F}(M)$, where the mass density $\rho$ is an element of $V^*$. Note
that in geophysical incompressible fluid dynamics there is also a second scalar
advected quantity, namely the buoyancy (for details, 
see  \cite{HoMaRa1998} and \cite{HoMaRa2002}) which plays the role that entropy
plays in a compressible fluid. In this case we would take $V= \mathcal{F}(M)
\times \mathcal{F}(M)$, where the second factor  is thought of as the space of
densities on $M$, thereby making the buoyancy, an element of its dual, into a
function.

\section{Hamiltonian formulation of Euler-Yang-Mills}\label{Hamiltonian}

Once  the Lagrangian formulation of a theory is known, one usually passes to
the
Hamiltonian formulation by a Legendre transformation, if the Lagrangian
function
is non-degenerate. Unfortunately, in our case, this is not possible because the
Legendre transformation is not invertible, as we have already seen when
studying
the Maxwell equations. The trouble is that the Lagrangian function does not
depend on
$\dot{\mathcal{A}}_0$. To deal with this, we shall work with a new Lagrangian
function obtained by eliminating $\mathcal{A}_0$ from \eqref{EYMLagrangian}.
For
this new Lagrangian function, the Legendre transformation is invertible and we
can deduce the associated Hamiltonian formulation. However, in this process, an
equation gets lost, namely, Gauss' Law $\delta^{ \mathcal{A}} \mathcal{E} = -
\mathcal{Q}$ in \eqref{EYM}. This equation will be recovered as a conservation
law of the momentum map associated to the gauge transformation group.
We begin by quickly recalling some facts about the Hamiltonian semidirect
product reduction theory.

\subsection{Hamiltonian semidirect product reduction with parameter} Let $S :=
G
\,\circledS\,V $ be the semidirect product defined at the beginning of
\S\ref{Lagrangian}. The lift of right translation of $S $ on $T ^\ast S $
induces a right action on $T ^\ast G \times V ^\ast$. Let $Q $ be another
manifold (without any $G $ or $V $-action). Consider a Hamiltonian function $H:
T ^\ast G \times T ^\ast Q \times V ^\ast \rightarrow \mathbb{R}$ right
invariant under the $S $-action on $T ^\ast G \times T ^\ast Q \times V ^\ast$;
recall that the $S $-action on $T ^\ast Q $ is trivial. In particular, the
function $H_{a_0}: = H|_{T ^\ast G \times T ^\ast Q \times \{a_0\}}: T ^\ast G
\times T ^\ast Q \rightarrow \mathbb{R}$ is invariant under the induced action
of the isotropy subgroup $G_{a_0}: = \{g \in G \mid \rho_g^\ast a_0 = a_0\}$
for
any $a_0 \in V ^\ast$. The following theorem is an easy consequence of the
semidirect product reduction theorem (see \cite{MaRaWe1984}) and the reduction
by stages method (see \cite{MaMiOrPeRa2007}).

\begin{theorem}\label{LPSD}
For $\alpha(t)\in T^*_{g(t)}G$ and
$\mu(t):=T^*R_{g(t)}(\alpha(t))\in\mathfrak{g}^*$, the following are
equivalent:
\begin{itemize}
\item[\bf{i}] $(\alpha(t),q(t),p(t))$ satisfies Hamilton's equations for
$H_{a_0}$ on $T^*(G\times Q)$.
\item[\bf{ii}] The following system of Lie-Poisson equations with parameter
coupled with Hamilton's equations holds on $\mathfrak{s}^*\times T^*Q$:
\[
\frac{\partial}{\partial t}(\mu,a)=-\operatorname{ad}^*_{\left(\frac{\delta
h}{\delta\mu},\frac{\delta h}{\delta a}\right)}(\mu,a)
=- \left(\operatorname{ad}^*_{\frac {\delta h}{ \delta \mu}}\mu+\frac {\delta
h}{ \delta a}\diamond
a,a\frac {\delta h}{ \delta \mu}\right),\quad a(0)=a_0
\]
and
\[
\frac{d q^i}{dt}=\frac{\partial h}{\partial p_i},\quad\frac{d
p_i}{dt}=-\frac{\partial h}{\partial q^i},
\]
where $\mathfrak{s}$ is the semidirect product Lie algebra
$\mathfrak{s}=\mathfrak{g}\,\circledS\, V$. The associated Poisson bracket is
the sum of the Lie-Poisson bracket on the Lie algebra $\mathfrak{s}^*$ and the
canonical bracket on the cotangent bundle $T^*Q$, that is,
\[
\{f,g\}(\mu,a,q,p)=\left\langle\mu,\left[\frac{\delta
f}{\delta\mu},\frac{\delta
g}{\delta\mu}\right]\right\rangle+\left\langle a,\frac{\delta f}{\delta
a}\frac{\delta g}{\delta\mu}-\frac{\delta g}{\delta a}\frac{\delta
f}{\delta\mu}\right\rangle+\frac{\partial f}{\partial q^i}\frac{\partial
g}{\partial p_i}-\frac{\partial g}{\partial q^i}\frac{\partial f}{\partial
p_i}.
\]
\end{itemize}
\end{theorem}

For example, one can start with a Lagrangian $L_{a_0}$ as in the previous
section, suppose that the Legendre transformation $\mathbb{F}L_{a_0}$ is
invertible, and form the corresponding Hamiltonian
$H_{a_0}=E_{a_0}\circ\mathbb{F}L_{a_0}^{-1}$, where $E_{a_0}$ is the energy of
$L _{a_0}$. Then the function $H: T^\ast G \times T ^\ast Q \times V ^\ast
\rightarrow \mathbb{R}$ so defined is $S $-invariant and one can apply this
theorem. This is the method we shall  use below to find the Hamiltonian
formulation of the Euler-Yang-Mills equations.

\subsection{The Hamiltonian} Recall that we identify the cotangent space
$T_\psi^*\mathcal{A}ut(P)$ with the space of $G$-invariant one-forms on $P$
along $\psi\in\mathcal{A}ut(P)$. The duality is given by
\[
\left\langle \mathbf{M}_\psi, U _\psi \right\rangle : = \int_M
\mathbf{M}_\psi(U
_\psi) \mu,
\]
where $\mathbf{M}_\psi \in T ^\ast_{\psi} \mathcal{A}ut (P)$ and $U _\psi \in
T_\psi \mathcal{A}ut (P)$. Recall that the integrand defines a function on $M $
(it is independent on the fiber variables in the bundle $\pi:P \rightarrow M
$).

For $U_\psi,V_\psi\in T_\psi\mathcal{A}ut(P)$, the expression
$K_\mathcal{A}(p)\left(U_\psi(p),V_\psi(p)\right)$ depends only on the class
$x=\pi(p)$. Thus $K_{\mathcal{A}}(U _\psi, V _\psi)$, which is a smooth
function
on $P $, does not depend on the fibers and hence induces a smooth function on
$M
$. Therefore the integral
\[
\int_MK_\mathcal{A}(U_\psi,V_\psi)\mu
\]
is well-defined. Moreover, the definition of $K _{ \mathcal{A}}$ immediately
implies the equality
\[
\int_MK_\mathcal{A}(U_\psi,V_\psi)\mu=\int_Mg([U_\psi],[V_\psi])\mu+\int_M\gamma\left(\widetilde{\mathcal{A}(U_\psi)},\widetilde{\mathcal{A}(V_\psi)}\right)\mu.
\]
Similarly, for $\mathbf{M}_\psi,\mathbf{N}_\psi\in T^*_\psi\mathcal{A}ut(P)$ we
can define the integral
\[
\int_MK^*_\mathcal{A}(\mathbf{M}_\psi,\mathbf{N}_\psi)\mu,
\]
where $K^*_\mathcal{A}$ denotes the dual metric induced on $T^*P$ by the
Kaluza-Klein metric.

The Hamiltonian for the Euler-Yang-Mills equations is defined on the cotangent
bundle $T^*(\mathcal{A}ut(P)\times\mathcal{C}onn(P))$ and is given, for
$(\rho,s)\in\mathcal{F}(M) \times \mathcal{F}(M)$, by
\begin{align}\label{EYMHamiltonian}
H_{(\rho,s)}(\mathbf{M}_\psi,\mathcal{A},\mathcal{Y})&=\frac{1}{2}\int_M\frac{1}{\rho}
K_\mathcal{A}^*(\mathbf{M}_\psi,\mathbf{M}_\psi)\mu+\int_M\rho
e(\rho(J\overline{\psi})^{-1},s)\mu\\
&\quad\quad+\frac{1}{2}\int_M\|E\|^2\mu+\frac{1}{2}\int_M\|B\|^2\mu.
\nonumber
\end{align}
This Hamiltonian is obtained by Legendre transforming the Lagrangian
$L_{(\rho,s)}$ in the case the variable $\mathcal{A}_0$ is absent. Indeed,
we have
\[
\mathbf{M}_\psi(p):=\mathbb{F}L(U_\psi)(p)=\rho(x)K_\mathcal{A}(p)(U_\psi(p),\cdot),\quad
x=\pi(p).
\]

By Theorem \ref{LPSD}, Hamilton's equations for $H_{(\rho,s)}$ are
equivalent to the Lie-Poisson equations on the dual of the semidirect product
Lie algebra $\mathfrak{aut}(P)\,\circledS\,(\mathcal{F}(M) \times
\mathcal{F}(M))$, together with the standard Hamilton equations on
$T^*\mathcal{C}onn(P)$, relative to the reduced Hamiltonian $h$ given on
$\big(\mathfrak{aut}(P)\,\circledS\,(\mathcal{F}(M) \times
\mathcal{F}(M))\big)^*\times T^*\mathcal{C}onn(P)$ 
by
\begin{align*} 
h(\mathbf{M},\rho,s,\mathcal{A},\mathcal{Y})& =\frac{1}{2}\int_M\frac{1}{\rho}K_\mathcal{A}^*(\mathbf{M},\mathbf{M})\mu+\int_M\rho
e(\rho,s)\mu \\
& \qquad +\frac{1}{2}\int_M\|E\|^2\mu+\frac{1}{2}\int_M\|B\|^2\mu.
\end{align*} 
By the Legendre transformation $U\mapsto \mathbf{M}=K_\mathcal{A}(\rho
U,\cdot)$, these equations are equivalent to the equations \eqref{EYM} with
$\mathcal{A}_0=0$ but without the nonabelian Gauss equation
\[
\quad\delta^\mathcal{A} \mathcal{E}=-\mathcal{Q}.
\]

\subsection{The momentum map of the gauge group} This last equation is obtained
by invariance of the Hamiltonian under gauge transformations. Indeed,
consider the action of the gauge group given for $\eta\in\mathcal{G}au(P)$, by
\begin{equation}
\label{gau_action}
(\psi,\mathcal{A})\mapsto(\eta^{-1}\circ\psi,\eta^*\mathcal{A}).
\end{equation}
The cotangent-lift of this action leaves the Hamiltonian invariant. So, the
associated momentum map, which is computed in the following lemma, is a
conserved quantity.

\begin{lemma} The momentum map associated to the cotangent lift of the gauge
group action is 
\[
\textbf{J}(\mathbf{M}_\psi,\mathcal{A},\mathcal{Y})=\sigma\left(\delta^\mathcal{A}\mathcal{Y}-\mathcal{A}\left(\left(J\overline{\psi}^{\,-1}
\right)V_\psi\circ\psi^{-1}\right)\right)\in\mathfrak{gau}(P)^* \simeq
\mathfrak{gau}(P),
\]
where $V_\psi\in T_\psi\mathcal{A}ut(P)$ is such that
$\mathbf{M}_\psi=K_\mathcal{A}(V_\psi,\cdot)$ and $\sigma: \mathcal{F}_G(P,
\mathfrak{g}) \rightarrow \mathfrak{gau}(P)$ is defined in \eqref{sigma}.
\end{lemma}
\textbf{Proof.} We will apply the formula
$\textbf{J}(\alpha_q)(\xi)=\langle\alpha_q,\xi_Q(q)\rangle$, which gives the
momentum mapping associated to a cotangent lifted action of a Lie group $G$ on
a
cotangent bundle $T^*Q$. In our case we have $G=\mathcal{G}au(P)$,
$Q=\mathcal{A}ut(P)\times\mathcal{C}onn(P)$ and for
$\xi=\sigma(f)\in\mathfrak{gau}(P)$, the infinitesimal generator is given by
(see \eqref{der_gau_connection})
\begin{align*}
\xi_Q(\psi,\mathcal{A})&=\left.\frac{d}{dt}\right|_{t=0}\left(\operatorname{exp}(t\xi)^{-1}\circ\psi,\operatorname{exp}(t\xi)^*\mathcal{A}\right)\\
&=\left(-\xi\circ\psi,\textbf{d}^\mathcal{A}f\right)\in
T_{(\psi,\mathcal{A})}(\mathcal{A}ut_\mu(P)\times\mathcal{C}onn(P)).
\end{align*}
Thus, changing variables in the third equality below, using \eqref{adjoint}, we
get
\begin{align*}
\langle\textbf{J}(\mathbf{M}_\psi&,\mathcal{A},\mathcal{Y}),\xi\rangle
=\langle(\mathbf{M}_\psi,\mathcal{A},\mathcal{Y}),(-\xi\circ\psi,\mathcal{A},\textbf{d}^\mathcal{A}f)\rangle\\
&=-\int_MK_\mathcal{A}(V_\psi,\xi\circ\psi)\mu+\int_M(g\gamma)\left(\widetilde{\mathcal{Y}},\widetilde{\textbf{d}^\mathcal{A}f}\right)\mu\\
&=-\int_MK_\mathcal{A}\left(\left(J\overline{\psi}^{\,-1}\right)V_\psi\circ\psi^{-1},\xi\right)\mu+\int_M\gamma\left(\widetilde{\delta^\mathcal{A}\mathcal{Y}},\widetilde{f}\right)\mu.
\end{align*}
Since $[\xi]=0$, the first term can be written as
\[
-\int_M\gamma\left(\widetilde{\mathcal{A}\left(\left(J\overline{\psi}^{\,-1}\right)V_\psi\circ\psi^{-1}\right)},\widetilde{\mathcal{A}(\xi)}\right)\mu.
\]
Thus, using the pairing \eqref{duality_for_Gau} and the identity
$\mathcal{A}\circ \sigma = id_{\mathcal{F}_G(P, \mathfrak{g})}$, we get
\begin{align*}
\langle\textbf{J}(\mathbf{M}_\psi&,\mathcal{A},\mathcal{Y}),\xi\rangle
=-\int_M\gamma\left(\widetilde{\mathcal{A}\left(\left(J\overline{\psi}^{\,-1}\right)V_\psi\circ\psi^{-1}\right)},\widetilde{\mathcal{A}(\xi)}\right)\mu \\
& \qquad \qquad \qquad +\int_M\gamma\left(\widetilde{\delta^\mathcal{A}\mathcal{Y}},\widetilde{\mathcal{A}(\xi)}\right)\mu\\
&=\left\langle\sigma\left(\delta^\mathcal{A}\mathcal{Y}-\mathcal{A}\left(\left(J\overline{\psi}^{\,-1}
\right)V_\psi\circ\psi^{-1}\right)\right),\xi\right\rangle.\qquad\blacksquare
\end{align*}

When $\mathbf{M}_\psi=\mathbb{F}L(U_\psi)=K_\mathcal{A}(\rho_0U_\psi,\cdot)$,
is
a solution of Hamilton's equations associated to $H_{(\rho_0,s_0)}$, the
conservation law $\textbf{J}(\mathbf{M}_\psi,\mathcal{A},\mathcal{Y})=0$ gives
\[
\mathcal{A}\left(\left(J\overline{\psi}^{\,-1} \right)(\rho_0\circ
\overline{\psi}^{\,-1})U_\psi\circ\psi^{-1}\right)=\delta^\mathcal{A}\mathcal{Y}.
\]
The definition of the charge density $\mathcal{Q}$ (see
\eqref{def_charge_density} without $\mathcal{A}_0$), the identities $U _\psi
\circ \psi^{-1} = U$, $ \left(J\overline{\psi}^{\,-1}
\right)(\rho_0\circ\overline{\psi}^{\,-1}) = \rho$,  and the  notation
$\mathcal{E}= - \mathcal{Y}$, gives
\[
\mathcal{Q}=-\delta^\mathcal{A}\mathcal{E}.
\]

The following theorem summarizes the results of the present section.
\begin{theorem} \label{hamiltonian_EYM_theorem}
Let $(\mathbf{M}_\psi,\mathcal{A},\mathcal{Y})$ be a curve in  $T^*(\mathcal{A}ut(P)\times\mathcal{C}onn(P))$ and consider the induced
curve $(\mathbf{M},\mathcal{A},\mathcal{Y})\in\mathfrak{aut}(P)^*\times
T^*\mathcal{C}onn(P)$ given by $\mathbf{M}: = \left(J \overline{ \psi} \right)
\mathbf{M}_ \psi \circ \psi^{-1}$. Then
$(\mathbf{M}_\psi,\mathcal{A},\mathcal{Y})$ is a solution of Hamilton's
equations associated to the Hamiltonian $H_{(\rho_0,s_0)}$ given in
\eqref{EYMHamiltonian} if and only if $(\mathbf{M},\mathcal{A},\mathcal{Y})$ is
a solution of the system
\begin{equation}
\label{EYM6}
\left\lbrace
\begin{array}{ll}
\displaystyle\frac{\partial v}{\partial
t}+\nabla_vv=\frac{1}{\rho}\gamma\left(Q,E(\cdot )+B( \cdot
,v)\right)^\sharp-\frac{1}{\rho}\operatorname{grad}p,\vspace{0.2cm}\\
\displaystyle\frac{\partial\rho}{\partial t}+\operatorname{div}(\rho
v)=0,\,\rho(0)=\rho_0,\quad\frac{\partial s}{\partial t}+{\bf d}s
(v)=0,\,s(0)=s_0,\vspace{0.2cm}\\
\displaystyle\frac{\partial Q}{\partial
t}+\nabla^\mathcal{A}_vQ+Q\operatorname{div}v=0,\vspace{0.2cm}\\
\displaystyle\displaystyle\frac{\partial \mathcal{E}}{\partial
t}=\delta^\mathcal{A}
\mathcal{B}-\mathcal{Q}\otimes\pi^*v^\flat,\vspace{0.2cm}\\
\displaystyle\frac{\partial \mathcal{B}}{\partial t}=-{\bf
d}^\mathcal{A}\mathcal{E}, \quad \mathbf{d}^ \mathcal{A}\mathcal{B} = 0,
\end{array} \right.
\end{equation}
where we use the same notations as in Theorem \ref{THM1} except that here $v$
and $\mathcal{Q}$ are given in terms of $\mathbf{M}$ by
\begin{equation}
\label{formulas_for_v_Q_U}
v=[U]\quad\text{and}\quad\mathcal{Q}=\mathcal{A}(\rho U)\quad\text{where}\quad
U=K_\mathcal{A}^*\left(\frac{\mathbf{M}}{\rho},\cdot\right).
\end{equation}
Conservation of the momentum map associated to the gauge transformations gives
the equation
\[
\quad\delta^\mathcal{A} \mathcal{E}=-\mathcal{Q}.
\]
\end{theorem}

One can adapt this theorem to the incompressible and homogeneous case.

\subsection{The Poisson bracket} From Theorem \ref{LPSD} we know that the
Euler-Yang-Mills equations \eqref{EYM6} can be written as
\[
\dot{f}=\{f,h\}
\]
with respect to the Poisson bracket on $[\mathfrak{aut}(P)
\,\circledS\,(\mathcal{F}(M) \times \mathcal{F}(M))]^\ast \times T ^\ast
\mathcal{C}onn(P)$
\begin{align}\label{Poisson_bracket}
\{f,g\}(\mathbf{M},\rho,&s,\mathcal{A},\mathcal{Y})=\int_M\mathbf{M}\left(\left[\frac{\delta
f}{\delta \mathbf{M}},\frac{\delta g}{\delta  \mathbf{M}}\right]_L\right)\mu\\
&+\int_M\rho\left(\textbf{d}\left(\frac{\delta f}{\delta
\rho}\right)\left[\frac{\delta g}{\delta 
\mathbf{M}}\right]-\textbf{d}\left(\frac{\delta g}{\delta
\rho}\right)\left[\frac{\delta f}{\delta  \mathbf{M}}\right]\right)\mu
\nonumber \\
&+\int_Ms\left(\operatorname{div}\left(\frac{\delta f}{\delta
s}\left[\frac{\delta g}{\delta 
\mathbf{M}}\right]\right)-\operatorname{div}\left(\frac{\delta g}{\delta
s}\left[\frac{\delta f}{\delta  \mathbf{M}}\right]\right)\right)\mu
\nonumber  \\
&+\int_M(g\gamma)\left(\frac{\delta f}{\delta \mathcal{A}},\frac{\delta
g}{\delta \mathcal{Y}}\right)\mu-\int_M(g\gamma)\left(\frac{\delta g}{\delta
\mathcal{A}},\frac{\delta f}{\delta \mathcal{Y}}\right)\mu.
\nonumber 
\end{align}

We can obtain this bracket and the associated Hamilton equations
\eqref{EYM6} alternatively by a reduction by stages process (see
\cite{MaMiOrPeRa2007}). The symplectic reduced spaces are of the form
$\mathcal{O} \times  T ^\ast \mathcal{C}onn(P)$, where  $\mathcal{O}$ is a coadjoint orbit of the semidirect product $S:=\mathcal{A}ut(P)
\,\circledS\,(\mathcal{F}(M) \times \mathcal{F}(M))$.

If the principal bundle is trivial, the automorphism group is the semidirect product
$\mathcal{D}(M) \,\circledS\, \mathcal{F}(M, G)$ of the
diffeomorphism group of $M$ with the group of $G $-valued functions on $M $. In
this case the first term can be written more explicitly by taking advantage of
the internal structure of $\mathcal{A}ut(P)$, and we recover (up to sign
conventions) the Poisson bracket given in equation (38) in \cite{GiHoKu1983}.

\subsection{Summary} We comment now on the structure of the equations of motion
\eqref{EYM6} and the Poisson bracket \eqref{Poisson_bracket}. Note that in
\eqref{EYM6} there is an evolution equation for the gauge charge $Q $ but that
the functions for which the Poisson bracket \eqref{Poisson_bracket} is defined
seem not to depend on $Q$. The explanation of this fact is given in Theorem 
\ref{hamiltonian_EYM_theorem}; the discussion below summarizes briefly the key
results and comments on the structure of both the equations and the Poisson
bracket.
\begin{itemize}
\item[(1)]  The equations for $v$ and $Q$ are the ``components" of a 
\textit{single\/} equation: the Lie- Poisson equation on the dual of the Lie
algebra of the automorphism group. The true variable is the fluid momentum
$\mathbf{M}$ which defines both the Eulerian velocity $v$ and the gauge-charge
$Q$ by using \eqref{formulas_for_v_Q_U}. Conversely, given $\rho$, $v$, and
$Q$,
the fluid momentum $\mathbf{M}$ is found by putting $\mathbf{M}: = \rho
K_\mathcal{A}(U,\cdot) = \rho
g(v,T\pi(\cdot))+\gamma(\mathcal{Q},\mathcal{A}(\cdot))$, where
$U:=\operatorname{Hor}_\mathcal{A}\circ v+\frac{1}{\rho}\sigma(\mathcal{Q})$
(recall that for any $U\in\mathfrak{aut}(P)$ and
$\mathcal{A}\in\mathcal{C}onn(P)$ we have the identity
$U=\operatorname{Hor}_\mathcal{A}\circ [U]+\sigma(\mathcal{A}(U))$). In other
words, the Lie-Poisson equation for $\mathbf{M}$ is equivalent to
\textit{two\/} equations: the equation for $v$ and the equation for $Q$. This
is
the usual Kaluza-Klein point of view, namely, $Q$ and $v$ are constructed from
$\mathbf{M}$ and vice-versa. 

\item[(2)]The $Q$-equation looks like advection equation.  To see this, recall
that  $Q \in \Gamma(\operatorname{Ad}P)$ and that
$\delta^\mathcal{A}:\Omega^1(M,\operatorname{Ad}P)\to\Gamma(\operatorname{Ad}P)$
(see Definition 4.2.8 in \cite{Bleecker1981}). Defining
\[
\operatorname{div}^\mathcal{A}(Qv):=-\delta^\mathcal{A}(Q\otimes
v^\flat),\qquad v \in \mathfrak{X}(M),
\]
where the one form $Q\otimes v^\flat\in\Omega^1(M,\operatorname{Ad}P)$ is given by 
$(Q\otimes v^\flat)(u_x):=Q(x)g_x(v(x),u_x)$, for any $u_x \in T_xM$,
one easily deduces the formula 
\[
\operatorname{div}^\mathcal{A}(Qv)=\nabla_v ^\mathcal{A} Q + Q
\operatorname{div} v
\]
which allows us to write the $Q $-equation in the form
\[
\dot{Q}+\operatorname{div}^\mathcal{A}(Qv)=0.
\]

However $Q $ is not advected since its evolution is not given by the pull back
of the flow of the velocity field $v $. Note that in this equation $\mathcal{A}$
is itself a variable that is time dependent.

\item[(3)] The equations for $\rho$ and $s$ are usual advection equations
for a density  and a function that appear in the Lie-Poisson equations of a
semidirect product.

\item[(4)] The equations for $\mathcal{E}$ and $\mathcal{B}$ are Hamilton's
equations for the conjugate variables $(\mathcal{A}, \mathcal{Y}) \in T ^\ast
\mathcal{C}onn(P)$.

\item[(5)] The equation $\delta^\mathcal{A}\mathcal{E}=-\mathcal{Q}$ comes from
momentum conservation associated to gauge group symmetry and
$\mathbf{d}^\mathcal{A} \mathcal{B} = 0$ is the Bianchi identity for the
connection
$\mathcal{A}$ and  its curvature $\mathcal{B}$.

\item[(6)] The Poisson bracket \eqref{Poisson_bracket} contains two types of
terms: the first three are Lie-Poisson for a semidirect product and the fourth
is the usual bracket on $T ^\ast \mathcal{C}onn(P)$. However, note that the
first summand in \eqref{Poisson_bracket} gives rise to an evolution equation
for
$\mathbf{M}$ which, as we discussed above, is equivalent to \textit{two\/}
evolution equations, one for $v $ and and  another one for $Q $. If the bundle
is trivial, one can make the formulas \eqref{formulas_for_v_Q_U} more explicit,
as we shall see below when we carry out one  more  reduction. Note also that
the
Poisson bracket \eqref{Poisson_bracket} is a product bracket; there is no
coupling between the semidirect product fluid variables $( \mathbf{M}, \rho,
s)$ and the Yang-Mills field variables $( \mathcal{A}, \mathcal{Y})$. The
coupling  in the equations is exclusively due to the Hamiltonian
\eqref{EYMHamiltonian}.
\end{itemize}

\subsection{The second reduction} Note that right translation in the group $S =
\mathcal{A}ut(P) \,\circledS\, (\mathcal{F}(M) \times \mathcal{F}(M))$ on
itself
and the action of $\mathcal{G}au(P)$ on $\mathcal{A}ut(P) \times
\mathcal{C}onn(P)$ given by \eqref{gau_action} commute if one views them as
actions on
$S\times \mathcal{C}onn(P)$. Therefore, by the general theory of commuting
reduction by stages (see \cite{MaMiOrPeRa2007}), since the momentum map
associated to
the gauge group action is $\mathcal{A}ut(P)$-invariant, it induces a momentum
map $\mathbf{J}_{\mathfrak{s}^*}$ on $\mathfrak{s}^*\times T ^\ast
\mathcal{C}onn(P)$ which restricts to a momentum map $\mathbf{J}_{\mathcal{O}}$
on the reduced space $\mathcal{O} \times  T ^\ast \mathcal{C}onn(P)$. Here $\mathfrak{s} : =
\mathfrak{aut}(P) \,\circledS\,(\mathcal{F}(M) \times \mathcal{F}(M))$.
A direct computation shows that the momentum map $\textbf{J}_{\mathfrak{s}^*} : \mathfrak{s}^* \times  T ^\ast
\mathcal{C}onn(P)\rightarrow
\mathfrak{gau}(P)^*$ has the expression
\begin{equation}\label{reduced_gauge_momentum}
\textbf{J}_{\mathfrak{s}^*}(\textbf{M},\rho,s,\mathcal{A},\mathcal{Y})=\sigma\left(\delta^\mathcal{A}\mathcal{Y}-\mathcal{A}(V)\right),
\end{equation}
where $V\in\mathfrak{aut}(P)$ is such that $\textbf{M}=K_\mathcal{A}(V,
\cdot)$,
and $\sigma$ denotes the map defined in \eqref{sigma}.
The gauge group action induced on $\mathfrak{s}^*\times T ^\ast
\mathcal{C}onn(P)$ and $\mathcal{O}\times T ^\ast \mathcal{C}onn(P)$ is given
by
\begin{equation}\label{reduced_gauge_action_0}
(\textbf{M},\rho,s,\mathcal{A},\mathcal{Y})\mapsto
(\operatorname{Ad}^*_\eta\textbf{M},\rho,s,\eta^*\mathcal{A},\eta^*\mathcal{Y}).
\end{equation}
Using the notation $\textbf{S}:=(\textbf{M},\rho,s)\in\mathcal{O}$, it can
be written as
\begin{equation}\label{reduced_gauge_action}
(\textbf{S},\mathcal{A},\mathcal{Y})\mapsto
(\operatorname{Ad}^*_{(\eta,0,0)}\textbf{S},\eta^*\mathcal{A},\eta^*\mathcal{Y}).
\end{equation}
This action is simply the diagonal action given on the first factor by the
coadjoint action of the subgroup $\mathcal{G}au(P)$ of
$S=\mathcal{A}ut(P)\,\circledS\,(\mathcal{F}(M) \times \mathcal{F}(M))$, and on
the second factor by gauge transformations. Note that when the center $Z(G)$
of the group $G$ is trivial, then the transformation
$\mathcal{A}\mapsto\eta^*\mathcal{A}$ is free. In this case, the reduced action
\eqref{reduced_gauge_action} is also free and the second reduced symplectic
spaces
\[
\textbf{J}^{-1}_{\mathcal{O}}(\textbf{N})/\mathcal{G}au(P)_{\textbf{N}},\quad
\textbf{N}\in\mathfrak{gau}(P)^*,
\]
have no singularities.

By the reduction by stages process, the reduced spaces
$\textbf{J}^{-1}_{\mathcal{O}}(\textbf{N})/\mathcal{G}au(P)_{\textbf{N}}$ are
symplectically diffeomorphic to the reduced spaces obtained by a one step
reduction from the cotangent bundle
\[
T^*\big(S\times\mathcal{C}onn(P)\big)
\]
with respect to the product of the two cotangent-lifted actions. Note that
these
reduced spaces are, up to connected components, the symplectic leaves in the
Poisson manifold
$\textbf{J}^{-1}_{\mathfrak{s}^*}(\mathbf{N})/\mathcal{G}au(P)_\mathbf{N}$.
This
is a straightforward consequence of Theorem 10.1.1(iv) in \cite{OrRa2004},
because the optimally reduced spaces are, up to connected components, precisely
the symplectically reduced spaces for every leaf. 

Note that the Euler-Yang-Mills
equation \eqref{EYM6} projects to the reduced space at zero momentum
\begin{equation}\label{reduction_at_zero}
\textbf{J}^{-1}_{\mathcal{O}}(0)/\mathcal{G}au(P).
\end{equation}
The general case corresponds to the Yang-Mills fluid with an external charge
$\mathbf{N}$.

In order to obtain the reduced Poisson structure concretely, we will identify
the space $\mathfrak{s}^*\times T^*\mathcal{C}onn(P)$ with a space on which the
gauge action is simpler. This identification is given in the following
proposition.

\begin{proposition}
Consider the group
$K=\mathcal{D}(M)\,\circledS\,(\mathcal{F}(M)\times\mathcal{F}(M))$ and denote
by $\mathfrak{k}^*$ the dual of its Lie algebra. There is a gauge-equivariant
diffeomorphism
\begin{equation}\label{bijection}
i:\mathfrak{s}^*\times T^*\mathcal{C}onn(P)\rightarrow\mathfrak{k}^*\times
\mathcal{F}_G(P,\mathfrak{g}^*)\times T^*\mathcal{C}onn(P),
\end{equation}
given by
\[
i(\mathbf{M},\rho,s,\mathcal{A},\mathcal{Y}):=
((\operatorname{Hor}_\mathcal{A})^*\circ\mathbf{M},\rho,s,\mathbb{J}\circ\mathbf{M},\mathcal{A},-\mathcal{Y})=:(\mathbf{n},\rho,s,\mathbf{\nu},\mathcal{A},\mathcal{E}),
\]
where the gauge group acts on $\mathfrak{s}^*\times T^*\mathcal{C}onn(P)$ by
the
action \eqref{reduced_gauge_action} and on $\mathfrak{k}^*\times
T^*\mathcal{C}onn(P)$ only on the factor $\mathcal{F}_G(P,\mathfrak{g}^*)\times
T^*\mathcal{C}onn(P)$ by the right action
\begin{equation}\label{gauge_action}
(\nu,\mathcal{A},\mathcal{E})\mapsto
(\nu\circ\eta,\eta^*\mathcal{A},\eta^*\mathcal{E}).
\end{equation}
Moreover, the image of the level set
$\mathbf{J}^{-1}_{\mathfrak{s}^*}(\mathbf{N})$ by the diffeomorphism $i$ is
\[
\{(\mathbf{n},\rho,s,\nu,\mathcal{A},\mathcal{E})\mid
\nu+\gamma(\delta^\mathcal{A}\mathcal{E}+f,\cdot)=0\},
\]
where $\mathbf{N}\in\mathfrak{gau}(P)$ and $f\in\mathcal{F}_G(P,\mathfrak{g})$
is such that $\sigma(f)=\mathbf{N}$. Thus
$\mathbf{J}^{-1}_{\mathfrak{s}^*}(\mathbf{N})$ is diffeomorphic to
$\mathfrak{k}^*\times T^*\mathcal{C}onn(P)$.

The map $\mathbb{J}:T^*P\rightarrow\mathfrak{g}^*$ denotes the momentum map
$\mathbb{J}(\alpha_p)(\xi):=\langle\alpha_p,\xi_P(p)\rangle$, and
$(\operatorname{Hor}_\mathcal{A})^*$ denotes the dual map of the
horizontal-lift
$\operatorname{Hor}_\mathcal{A}: TM \rightarrow TP$ with respect to
$\mathcal{A}$.
\end{proposition}
\textbf{Proof.}
We first prove that $i$ is injective. Suppose that
$(\mathbf{M},\rho,s,\mathcal{A},\mathcal{Y})$,
$(\mathbf{M}',\rho',s',\mathcal{A}',\mathcal{Y}')\in\mathfrak{s}^*\times
T^*\mathcal{C}onn(P)$ have the same image under $i$. We clearly have
$(\rho,s,\mathcal{A},\mathcal{Y})=(\rho',s',\mathcal{A}',\mathcal{Y}')$.
Therefore we have 
$(\operatorname{Hor}_\mathcal{A})^*\circ\mathbf{M}=(\operatorname{Hor}_\mathcal{A})^*\circ\mathbf{M}'$
and $\mathbb{J}\circ\mathbf{M}=\mathbb{J}\circ\mathbf{M}'$. This implies that
$\mathbf{M}=\mathbf{M}'$. The map $i$ is clearly surjective and hence
invertible, its inverse being given by
\[
i^{-1}(\mathbf{n},\rho,s,\nu,\mathcal{A},\mathcal{E})=(\pi^*\mathbf{n}+\mathcal{A}^*\nu,\rho,s,\mathcal{A},-\mathcal{E}),
\]
where $\mathcal{A}^* :\mathfrak{g}^*\rightarrow T^*P$ denotes the dual map of
$\mathcal{A}$. It follows that $i$ is a diffeomorphism.

To prove gauge-equivariance, it suffices to show that for all
$\eta\in\mathcal{G}au(P)$,
\[
(\operatorname{Hor}_\mathcal{\eta^*A})^*\circ\eta^*\mathbf{M}=(\operatorname{Hor}_\mathcal{A})^*\circ\mathbf{M}\quad\text{and}\quad\mathbb{J}\circ\eta^*\mathbf{M}=(\mathbb{J}\circ\mathbf{M})\circ\eta.
\]
This is a direct computation using the formulas
\[
T^*_p\pi(\operatorname{Hor}_\mathcal{A})^*_p(\mathbf{M}(p))=\mathbf{M}(p)-\mathcal{A}(p)^*(\mathbb{J}(\mathbf{M}(p)))
\]\[
\eta^*\mathcal{A}=\operatorname{Ad}_{\widehat{\eta}^{-1}}\circ
\mathcal{A}+TL_{\widehat{\eta}^{-1}}\circ T\widehat{\eta}
\]
\[
\eta^*\mathbf{M}(p)=\mathbf{M}(p)+T^*_p\widehat{\eta}(T^*R_{\widehat{\eta}(p)^{-1}}(\mathbb{J}(\mathbf{M}(p)))),
\]
where $\widehat{\eta}\in\mathcal{F}_G(P,G)$ is such that
$\eta(p)=\Phi_{\widehat{\eta}(p)}(p)$.

Recall that
$\mathbf{J}_{\mathfrak{s}^*}(\mathbf{M},\rho,s,\mathcal{A},\mathcal{Y})=\sigma(\delta^\mathcal{A}\mathcal{Y}-\mathcal{A}(V))$,
where $V\in\mathfrak{aut}(P)$ is such that $\mathbf{M}=K_\mathcal{A}(V,\cdot)$.
So for $\mathbf{N}=\sigma(f)$ the condition
$\mathbf{J}_{\mathfrak{s}^*}(\mathbf{M},\rho,s,\mathcal{A},\mathcal{Y})=\mathbf{N}$
reads $\delta^\mathcal{A}\mathcal{Y}-\mathcal{A}(V)=f$. Using that
$\nu=\mathbb{J}\circ\mathbf{M}=\gamma(\mathcal{A}(V),\cdot)$ and
$\mathcal{E}=-\mathcal{Y}$, we get the condition
\[
\nu+\gamma(\delta^\mathcal{A}\mathcal{E}+f,\cdot)=0.\qquad\blacksquare
\]
\medskip

This proposition shows that the reduced spaces
$\mathbf{J}^{-1}_{\mathfrak{s}^*}(\mathbf{N})/\mathcal{G}au(P)_\mathbf{N}$ can
be identified with the quotient
$\mathfrak{k}^*\times\left[\left(\mathcal{F}_G(P,\mathfrak{g}^*)\times
T^*\mathcal{C}onn(P)\right)/\mathcal{G}au(P)_\mathbf{N}\right]$ via the
diffeomorphism induced by $i$ and given by
\begin{equation}\label{induced_i}
[(\mathbf{M},\rho,s,\nu,\mathcal{A},\mathcal{Y})]\mapsto
((\operatorname{Hor}_\mathcal{A})^*\circ\mathbf{M},\rho,s,[\nu,\mathcal{A},\mathcal{E}]),
\end{equation}
where $[\cdot]$ denote the corresponding equivalence classes.
\medskip

We now compute the Poisson structure $\{\,,\}'$ induced by $i$ on
$\mathfrak{k}^*\times\mathcal{F}_G(P,\mathfrak{g}^*)\times
T^*\mathcal{C}onn(P)$. For
$f,g\in\mathcal{F}(\mathfrak{k}^*\times\mathcal{F}_G(P,\mathfrak{g}^*)\times
T^*\mathcal{C}onn(P))$ we have the formulas
\[
\frac{\delta (f\circ i)}{\delta
\mathbf{M}}=\operatorname{Hor}_\mathcal{A}\circ\frac{\delta f}{\delta
\mathbf{n}}+\mathbb{J}^*\circ\frac{\delta f}{\delta
\mathbf{\nu}},\qquad \left[\frac{\delta (f\circ i)}{\delta
\mathbf{M}}\right]=\frac{\delta f}{\delta
\mathbf{n}},\qquad\mathbf{n}:=(\operatorname{Hor}_\mathcal{A})^*\circ\mathbf{M},
\]
\[
(g\gamma)\left(\frac{\delta (f\circ i)}{\delta
\mathcal{A}},\mathcal{C}\right)=(g\gamma)\left(\frac{\delta f}{\delta
\mathcal{A}},\mathcal{C}\right)-\nu\left(\mathcal{C}\left(\operatorname{Hor}_\mathcal{A}\left(\frac{\delta
f}{\delta \mathbf{n}}\right)\right)\right),\quad\nu:=\mathbf{J}\circ\mathbf{M}.
\]
Using the equality \eqref{bigformula}, we obtain
\begin{align}\label{Poisson_bracket'}
\{f,g&\}'(\mathbf{n},\rho,s,\nu,\mathcal{A},\mathcal{E}):
=\{f\circ i,g\circ i\}(\mathbf{M},\rho,s,\mathcal{A},\mathcal{Y})\\
&=\int_M\mathbf{n}\left(\left[\frac{\delta f}{\delta
\mathbf{n}},\frac{\delta g}{\delta
\mathbf{n}}\right]_L\right)\mu+\int_M\rho\left(\textbf{d}\left(\frac{\delta
f}{\delta \rho}\right)\frac{\delta g}{\delta
\mathbf{n}}-\textbf{d}\left(\frac{\delta g}{\delta \rho}\right)\frac{\delta
f}{\delta \mathbf{n}}\right)\mu  \nonumber  \\
&+\int_Ms\left(\operatorname{div}\left(\frac{\delta f}{\delta
s}\frac{\delta g}{\delta
\mathbf{n}}\right)-\operatorname{div}\left(\frac{\delta g}{\delta
s}\frac{\delta f}{\delta \mathbf{n}}\right)\right)\mu \nonumber  \\
&+\int_M(g\gamma)\left(\frac{\delta g}{\delta \mathcal{A}},\frac{\delta
f}{\delta \mathcal{E}}\right)\mu-\int_M(g\gamma)\left(\frac{\delta f}{\delta
\mathcal{A}},\frac{\delta g}{\delta \mathcal{E}}\right)\mu \nonumber  \\
&+\int_M\nu\left(\left[\frac{\delta f}{\delta\nu},\frac{\delta
g}{\delta\nu}\right] \right) \mu \nonumber \\
&+\int_M\nu\left(\frac{\delta g}{\delta
\mathcal{E}}\left(\operatorname{Hor}_\mathcal{A}\circ\frac{\delta f}{\delta
\mathbf{n}}\right)
-\frac{\delta f}{\delta
\mathcal{E}}\left(\operatorname{Hor}_\mathcal{A}\circ\frac{\delta g}{\delta
\mathbf{n}}\right)\right. \nonumber \\
&\qquad\qquad+\mathbf{d}^\mathcal{A}\left(\frac{\delta
f}{\delta\nu}\right)\left(\operatorname{Hor}_\mathcal{A}\circ\frac{\delta
g}{\delta \mathbf{n}}\right)
-\mathbf{d}^\mathcal{A}\left(\frac{\delta
g}{\delta\nu}\right)\left(\operatorname{Hor}_\mathcal{A}\circ\frac{\delta
f}{\delta \mathbf{n}}\right) \nonumber  \\
&\qquad\qquad\left.+\mathcal{B}\left(\operatorname{Hor}_\mathcal{A}\circ\frac{\delta
f}{\delta \mathbf{n}},\operatorname{Hor}_\mathcal{A}\circ\frac{\delta g}{\delta
\mathbf{n}}\right)\right)\mu. \nonumber 
\end{align}

Note that the first three terms in \eqref{Poisson_bracket'} represent the
Lie-Poisson bracket on $\mathfrak{k}^*$, the fourth and fifth terms represent
the canonical bracket on $T^*\mathcal{C}onn(P)$, the sixth term is the
Lie-Poisson bracket on $\mathcal{F}_G(P, \mathfrak{g}^\ast)$, and the last term
provides the coupling of the fluid variables to the Yang-Mills fields.
\medskip

By the general process of Poisson (point) reduction, the reduced spaces 
\[
\mathbf{J}_{\mathfrak{s}^*}^{-1}(\mathbf{N})/\mathcal{G}au(P)_\mathbf{N}\simeq\mathfrak{k}^*\times\left[\left(\mathcal{F}_G(P,\mathfrak{g}^*)\times
T^*\mathcal{C}onn(P)\right)/\mathcal{G}au(P)_\mathbf{N}\right]
\]
inherit a Poisson bracket  $\{\,,\}_\mathbf{N}$ given by
\begin{equation}\label{Reduced_Poisson_Bracket}
\{f_\mathbf{N},g_\mathbf{N}\}_\mathbf{N}(\mathbf{n},\rho,s,[\nu,\mathcal{A},\mathcal{E}]):=\{f,g\}'(\mathbf{n},\rho,s,\nu,\mathcal{A},\mathcal{Y}),
\end{equation}
where $f,g$ are any $\mathcal{G}au(P)$-invariant extensions of the functions
$f_\mathbf{N}\circ\pi_\mathbf{N}$, $g_\mathbf{N}\circ\pi_\mathbf{N}:\mathbf{J}_{\mathfrak{s}^*}^{-1}(\mathbf{N})\rightarrow\mathbb{R}$,
relative to the projection $\pi_\mathbf{N} :
\mathbf{J}^{-1}_{\mathfrak{s}^*}(\mathbf{N})\rightarrow\mathbf{J}_{\mathfrak{s}^*}^{-1}(\mathbf{N})/\mathcal{G}au(P)_\mathbf{N}$.

There  are no explicit formulas for the equations of motion on the Poisson point reduced space $\textbf{J}^{-1}_{\mathcal{O}}(0)/\mathcal{G}au(P) $
 because there is no concrete realization of this quotient, to our
knowledge. However, there is  an important particular case where this is
possible that we study next.

\subsection{The case of a trivial bundle} We end this section by examining the
case of a trivial principal bundle
$P=M\times G$ and, more precisely, the case of the Euler-Maxwell equations
which
are obtained by taking
by $G=S^1$. We then compare our results to those obtained for Euler-Maxwell in
\cite{MaWeRaScSp1983}. Recall that we have
$\mathfrak{aut}(P)=\mathfrak{X}(M)\,\circledS\,\mathcal{F}(M,\mathfrak{g})$, so we obtain
$\mathfrak{aut}(P)^*=\mathfrak{X}(M)^*\,\circledS\,\mathcal{F}(M,\mathfrak{g})^*=\Omega^1(M)\,\circledS\,\mathcal{F}(M,\mathfrak{g}^*)$.
For $(\mathbf{m},\nu)\in\mathfrak{aut}(P)^*$, the gauge transformation
\eqref{reduced_gauge_action_0} is given by
\begin{align*} 
&(\mathbf{m},\nu,\rho,s,A,Y)\mapsto \\
& \quad  (\mathbf{m}+T^*\widehat{\eta}\circ
T^*R_{\widehat{\eta}^{-1}}\circ\nu,\operatorname{Ad}^*_{\widehat{\eta}}\circ\nu,\rho,s,\operatorname{Ad}_{\widehat{\eta}^{-1}}\circ
A+TL_{\widehat{\eta}^{-1}}\circ
T\widehat{\eta},\operatorname{Ad}_{\widehat{\eta}^{-1}}\circ Y),
\end{align*} 
where $A:=\overline{\mathcal{A}},
Y:=\overline{\mathcal{Y}}\in\Omega(M,\mathfrak{g})$. The equivariant
diffeomorphism \eqref{bijection} is 
\[
i(\mathbf{m},\nu,\rho,s,A,Y)=(\mathbf{m}-A^*\nu,\rho,s,\nu,A,-Y)=:(\mathbf{n},\rho,s,\nu,A,E),
\]
and the gauge transformation \eqref{gauge_action} is
\begin{equation}\label{gauge_action_trivial}
(\mathbf{n},\rho,s,\nu,A,E)\mapsto(\mathbf{n},\rho,s,\operatorname{Ad}_{\widehat{\eta}}^*\circ\nu,\operatorname{Ad}_{\widehat{\eta}^{-1}}\circ
A+TL_{\widehat{\eta}^{-1}}\circ
T\widehat{\eta},\operatorname{Ad}_{\widehat{\eta}^{-1}}\circ E).
\end{equation}
Recall that the relation between the charge density $\mathcal{Q}$ and the
momentum $\mathbf{M}$ is
\[
\mathcal{Q}=\mathcal{A}(V),
\]
where $V\in\mathfrak{aut}(P)$ is such that $\mathbf{M}=K_\mathcal{A}(V,\cdot)$;
see \eqref{formulas_for_v_Q_U}. When the bundle is trivial, this relation reads
\begin{equation}\label{charge}
\nu=\gamma(Q,\cdot).
\end{equation}

In the case of Euler-Maxwell, since $G=S^1$, the gauge transformation is simply
\[
(\mathbf{m},\nu,\rho,s,A,Y)\mapsto
(\mathbf{m}+\nu\textbf{d}\eta,\nu,\rho,s,A+\textbf{d}\eta,Y),
\]
and the relation \eqref{charge} reads $\nu=Q$. Recall that we can write
$Q=\rho\frac{q}{m}$, where $q\in\mathcal{F}(M)$ is the charge, see Corollary
\ref{corollary_EM}. This gauge transformation coincides with the one given in
equation $(36)$ in \cite{MaWeRaScSp1983}, where the notation $a:=\frac{q}{m}$
is
used. The zero level set of the momentum map is 
\[
\mathbf{J}_{\mathfrak{s}^*}^{-1}(0)=\left\{\left(\mathbf{m},\rho\frac{q}{m},\rho,s,A,Y\right)\,
\left|  \, \delta Y=\rho\frac{q}{m},\right. \right\}.
\]
The bijection $i$ reads
\[
i\left(\mathbf{m},\rho\frac{q}{m},\rho,s,A,Y\right)=\left(\mathbf{m}-A\rho\frac{q}{m},\rho,s,\rho\frac{q}{m},A,-Y\right),
\]
and the image of $\mathbf{J}_{\mathfrak{s}^*}^{-1}(0)$ is 
\[
\{(\mathbf{n},\rho,s,\nu,A,E)\mid \operatorname{div}\mathbf{E}=\nu\},
\]
where the notation $\mathbf{E}:=E^\sharp\in\mathfrak{X}(M)$ is used. The gauge
transformation \eqref{gauge_action} is 
\[
(\mathbf{n},\rho,s,\nu,A,E)\mapsto(\mathbf{n},\rho,s,\nu,A+\mathbf{d}\eta,E).
\]
Through the diffeomorphism $i$, the projection $\pi_0
:\mathbf{J}_{\mathfrak{s}^*}^{-1}(0)\rightarrow\mathbf{J}_{\mathfrak{s}^*}^{-1}(0)/\mathcal{G}au(P)$
is given by
\[
(\mathbf{n},\rho,s,\nu,A,E)\mapsto(\mathbf{n},\rho,s,[A],E),
\]
where $[A]$ denotes the cohomology class of the one-form $A$. Assuming that the
first and second cohomology groups of $M$ are trivial, $H^1(M)=H^2(M)=\{0\}$,
we
get  the isomorphism
\begin{equation}\label{isomorphism}
[A]\mapsto B:=\mathbf{d}A\in\Omega^2_{cl}(M),
\end{equation}
where $\Omega^2_{cl}(M)$ denotes the space closed two-forms. Thus $i$ induces a
diffeomorphism between $\mathbf{J}_{\mathfrak{s}^*}^{-1}(0)/\mathcal{G}au(P)$
and the space $\mathfrak{k}^*\times\Omega^2_{cl}(M)\times\Omega^1(M)$ given by
\[
[\mathbf{m},\nu,\rho,s,A,Y]\mapsto
(\mathbf{m}-A\nu,\rho,s,\mathbf{d}A,-Y)=:(\mathbf{n},\rho,s,B,E).
\]
where $\Omega^2_{cl}(M)$ denotes the closed $2$-forms on $M$. This
identification coincides with the one given in Proposition 10.1 in
\cite{MaWeRaScSp1983}. 

Using the definition \eqref{Reduced_Poisson_Bracket} and the bracket
\eqref{Poisson_bracket'}, the reduced Poisson bracket on
$\mathfrak{k}^*\times\Omega^2_{cl}(M)\times\Omega^1(M)$ is 
\begin{align}\label{Maxwell_Poisson_bracket''}
\{f,g\}_0(\mathbf{n},&\rho,s,B,E)=\int_M\mathbf{n}\left(\left[\frac{\delta
f}{\delta \mathbf{n}},\frac{\delta g}{\delta \mathbf{n}}\right]_L\right)\mu\\
&+\int_M\rho\left(\textbf{d}\left(\frac{\delta f}{\delta
\rho}\right)\frac{\delta g}{\delta \mathbf{n}}-\textbf{d}\left(\frac{\delta
g}{\delta \rho}\right)\frac{\delta f}{\delta \mathbf{n}}\right)\mu \nonumber \\
&+\int_Ms\left(\operatorname{div}\left(\frac{\delta f}{\delta
s}\frac{\delta g}{\delta
\mathbf{n}}\right)-\operatorname{div}\left(\frac{\delta g}{\delta
s}\frac{\delta f}{\delta \mathbf{n}}\right)\right)\mu \nonumber  \\
&+\int_Mg\left(\delta\frac{\delta g}{\delta B},\frac{\delta f}{\delta
E}\right)\mu-\int_Mg\left(\delta\frac{\delta f}{\delta B},\frac{\delta
g}{\delta E}\right)\mu \nonumber  \\
&+\int_M\rho\frac{q}{m}\left(\frac{\delta g}{\delta E}\left(\frac{\delta
f}{\delta \mathbf{n}}\right)-\frac{\delta f}{\delta E}\left(\frac{\delta
g}{\delta \mathbf{n}}\right)+B\left(\frac{\delta f}{\delta
\mathbf{n}},\frac{\delta g}{\delta \mathbf{n}}\right)\right)\mu, \nonumber 
\end{align}
and the Euler-Maxwell equations can be written as
\[
\dot{f}=\{f,h\}_0,
\]
relative to the induced Hamiltonian $h$ given by
\begin{align*} 
h(\mathbf{n},\rho,s,B,E)=&\frac{1}{2}\int_M\left(\frac{1}{\rho}g(\mathbf{n},\mathbf{n})+\frac{1}{\rho}(\delta
E)^2\right)\mu+\int_M\rho
e(\rho,s)\mu \\
& \qquad +\frac{1}{2}\int_M\left(\|E\|^2+\|B\|^2\right)\mu.
\end{align*} 
Note that the function
\[
C(\mathbf{n},\rho,s,B,E)=\frac{1}{2}\int_M\frac{1}{\rho}(\delta E)^2
\]
is a Casimir function, so an equivalent Hamiltonian is 
\[
\overline{h}(\mathbf{n},\rho,s,B,E)=\frac{1}{2}\int_M\frac{1}{\rho}g(\mathbf{n},\mathbf{n})\mu+\int_M\rho
e(\rho,s)\mu+\frac{1}{2}\int_M\left(\|E\|^2+\|B\|^2\right)\mu.
\]
When $M$ is three dimensional, we can use the notations $\mathbf{B}:=(\star
B)^\sharp$ and $\mathbf{E}:=E^\sharp$. Therefore the two last terms can be
written as
\begin{align}
&\int_Mg\left(\operatorname{curl}\frac{\delta g}{\delta
\mathbf{B}},\frac{\delta f}{\delta
\mathbf{E}}\right)\mu-\int_Mg\left(\operatorname{curl}\frac{\delta f}{\delta
\mathbf{B}},\frac{\delta g}{\delta \mathbf{E}}\right)\mu\nonumber\\
&\quad+\int_M\rho\frac{q}{m}\left(g\left(\frac{\delta g}{\delta
\mathbf{E}},\frac{\delta f}{\delta \mathbf{n}}\right)-g\left(\frac{\delta
f}{\delta \mathbf{E}},\frac{\delta g}{\delta
\mathbf{n}}\right)+g\left(\mathbf{B},\frac{\delta f}{\delta
\mathbf{n}}\times\frac{\delta g}{\delta \mathbf{n}}\right)\right)\mu.\nonumber
\end{align}

This bracket coincides with the one derived in \cite{MaWeRaScSp1983} by a
direct
computation. Note that the first line in the formula above is the
Pauli-Born-Infeld Poisson bracket for the Maxwell equations (see, e.g.
\cite{MaRa1994}, \S1.6). The Hamiltonian $\bar{h}$ is very simple: it is the
sum
of the total energy of the fluid plus the energy of the electromagnetic field.

\medskip \noindent{\bf Remark.}
In the Euler-Maxwell case, the correspondence \eqref{isomorphism} is a
bijective
map if $H^1(M)=H^2(M)=\{0\}$. Indeed, for $B,B'$ such that
$\textbf{d}B=\textbf{d}B'=0$, we have $B=\textbf{d}A$ and $B'=\textbf{d}A'$,
therefore if $B=B'$ we have $A=A'+\textbf{d}\eta$, that is, $[A]=[A']$.

This fact does not generalize to the case of a nonabelian principal bundle,
trivial or not: there exist gauge inequivalent connections  (even on
$\mathbb{R}^3 \times G$) with the same curvatures and holonomy groups; see
\cite{Mo1986}, \cite{GuYa1977}, \cite{Mo1980}.

\section{The Kelvin-Noether Theorem}
\label{section:KN}

The Kelvin-Noether theorem is a version of the Noether theorem that holds for
solutions of the Euler-Poincar\'e equations. An application of this theorem to
the ideal compressible adiabatic fluid (see \eqref{ICAF}) gives the Kelvin
circulation theorem
\[
\frac{d}{dt}\oint_{\gamma_t}v^\flat=\oint_{\gamma_t}T\textbf{d}s,
\]
where $\gamma_t\subset M$ is a closed curve which moves with the fluid velocity
$v$, and $T=\partial e/\partial s$ is the temperature.

\subsection{Kelvin-Noether Theorem for semidirect products} In order to apply
this theorem to the Yang-Mills fluid, we recall some facts about the
Kelvin-Noether theorem for semidirect products (see \cite{HoMaRa1998} for
details).

We start with a Lagrangian $L_{a_0}$ depending on a parameter $a_0\in V^\ast$,
as at the beginning of \S\ref{Lagrangian}. We introduce a manifold
$\mathcal{C}$
on which $G$ acts on the left and suppose we have an equivariant map
$\mathcal{K}:\mathcal{C}\times V^*\rightarrow\mathfrak{g}^{**}$, that is, for
all $g\in G, a\in V^*, c\in\mathcal{C}$, we have
\[
\langle\mathcal{K}(gc,\rho^*_g(a)),\mu\rangle=\langle\mathcal{K}(c,a),\operatorname{Ad}_g^*\mu\rangle,
\]
where $gc$ denotes the action of $G$ on $\mathcal{C}$.

Define the Kelvin-Noether quantity $I:\mathcal{C}\times\mathfrak{g}\times
TQ\times V^*\rightarrow\mathbb{R}$ by
\[
I(c,\xi,q,\dot{q},a):=\left\langle\mathcal{K}(c,a),\frac{\delta
l}{\delta\xi}(\xi,q,\dot{q},a)\right\rangle.
\]

\begin{theorem} {\bf (Kelvin-Noether.)}\label{KN}
Fixing $c_0\in\mathcal{C}$, let $\xi(t),q(t),\dot{q}(t),a(t)$ satisfy the
Euler-Poincar\'e equations and define $g(t)$ to be the solution of
$\dot{g}(t)=TR_{g(t)}\xi(t)$ and, say, $g(0)=e$. Let $c(t)=g(t)c_0$ and
$I(t):=I(c(t),\xi(t),q(t),\dot{q}(t),a(t))$. Then
\[
\frac{d}{dt}I(t)=\left\langle\mathcal{K}(c(t),a(t)),\frac{\delta l}{\delta
a}\diamond a\right\rangle.
\]
\end{theorem}

\subsection{The Kelvin-Noether Theorem for Yang-Mills fluids} In the case of
the
Yang-Mills fluid, we shall choose for the abstract Lie group $G$ above, the
automorphism group $\mathcal{A}ut(P)$ and we let
$\mathcal{C}=\{c\in\mathcal{F}(S^1,P)\mid \pi\circ c\in
\operatorname{Emb}(S^1,M)\}$, where $\operatorname{Emb}(S^1,M)$ denotes the
manifold of all embeddings of the circle $S^1$ in $M$. The left action of
$\mathcal{A}ut(P)$ on $\mathcal{C}$ is given by $c\mapsto\varphi\circ c$. The
map $\mathcal{K}$ is defined by
\[
\left\langle\mathcal{K}(c,(\rho,s)),\mathbf{M}\right\rangle:=\oint_c\frac{1}{\rho\circ\pi}\mathbf{M},
\qquad \mathbf{M} \in \Omega^1_G(P).
\]
A change of variables in the integral shows that $\mathcal{K}$ is equivariant,
that is,
\[
\langle\mathcal{K}(\varphi\circ
c,(J\overline{\varphi}^{-1}(\rho\circ\overline{\varphi}^{-1}),s\circ\overline{\varphi}^{-1}),\mathbf{M}\rangle=\langle\mathcal{K}(c,(\rho,s)),\operatorname{Ad}^*_{\varphi}\mathbf{M}\rangle.
\]

Using the Lagrangian $l$ given in \eqref{reducedEYMLagrangian}, we have
\[
\frac{\delta l}{\delta
U}(U,\mathcal{A}_0,\dot{\mathcal{A}}_0,\mathcal{A},\dot{\mathcal{A}},(\rho,s))=\pi^*(\rho
g([U],\cdot))+\gamma(\mathcal{Q},\mathcal{A}(\cdot)).
\]
Therefore the Kelvin-Noether quantity is
\[
I(c,U,\mathcal{A}_0,\dot{\mathcal{A}}_0,\mathcal{A},\dot{\mathcal{A}},(\rho,s))=\oint_{\overline{c}}[U]^\flat+\oint_c\frac{1}{\rho\circ\pi}\gamma\left(\mathcal{Q},\mathcal{A}(\cdot)\right),
\]
where $\overline{c}:=\pi\circ c\in\operatorname{Emb}(S^1,M)$.

On the other hand, using the equality
\[
\frac{\delta l}{\delta (\rho,s)}\diamond
(\rho,s)=T^*\pi\left(\rho\,\textbf{d}\frac{\delta
l}{\delta\rho}-\frac{\delta l}{\delta s}\textbf{d}s\right),
\]
we get
\begin{align*}
\left\langle\mathcal{K}(c,(\rho,s)),\frac{\delta l}{\delta
(\rho,s)}\diamond
(\rho,s)\right\rangle=\oint_{\overline{c}}\frac{\partial
e}{\partial s}\textbf{d}s.
\end{align*}
Thus, by Theorem \ref{KN}, the \textit{Kelvin Circulation Theorem for the
Yang-Mills fluid\/} is
\begin{equation}\label{KNEYM}
\frac{d}{dt}\left[\oint_{\overline{\gamma}_t}v^\flat+\oint_{\gamma_t}\frac{1}{\rho\circ\pi}\gamma\left(\mathcal{Q},\mathcal{A}(\cdot)\right)\right]=\oint_{\overline{\gamma}_t}T\textbf{d}s,
\end{equation}
where $\gamma_t:=\varphi_t\circ c_0\subset P$,
$\overline{\gamma}_t=\overline{\varphi_t}\circ \overline{c}_0\subset M$ is a
closed curve which moves with the fluid velocity $v:=[U]$, and $T:=\partial
e/\partial s$ is the temperature.

When the principal bundle is trivial, formula \eqref{KNEYM} reads
\[
\frac{d}{dt}\left[\oint_{\overline{\gamma}_t}
\left(v^\flat+\frac{1}{\rho}\gamma\left(Q,\overline{\mathcal{A}}(\cdot)\right)
\right)+\oint_{\gamma_t}\frac{1}{\rho}\gamma(Q,\cdot)\right]
=\oint_{\overline{\gamma}_t}T\textbf{d}s.
\]
For the Euler-Maxwell fluid consisting of particles of mass $m $ and charge $q
$, since $Q=\rho\frac{q}{m}$, the second integral vanishes and we get
\[
\frac{d}{dt}\oint_{\overline{\gamma}_t}\left( v^\flat+\frac{q}{m}A \right)
=\oint_{\overline{\gamma}_t}T\textbf{d}s,
\]
which coincides with formula (7.37) in \cite{HoMaRa1998}.

\medskip

\noindent \textbf{Acknowledgments.} We thank Jerry Marsden for drawing our
attention to this problem and for many illuminating  discussions. Our thanks go
to Marco Castrill\'on-L\'opez, Darryl Holm, Juan-Pablo Ortega, and the anonymous referee for several remarks that improved our exposition.

{\footnotesize

\bibliographystyle{new}

}

\end{document}